\documentclass[journal,twoside,web]{ieeecolor}
\usepackage{generic}
\usepackage{cite}
\usepackage{amsmath,amssymb,amsfonts}
\usepackage{algorithmic}
\usepackage{subfig}
\usepackage{graphicx}
\usepackage{textcomp}
\def\BibTeX{{\rm B\kern-.05em{\sc i\kern-.025em b}\kern-.08em
    T\kern-.1667em\lower.7ex\hbox{E}\kern-.125emX}}
 \usepackage{epsfig}
 \usepackage[latin1]{inputenc}
\usepackage{graphicx}      
\usepackage{amsmath,amssymb}
\usepackage{pstricks,pst-node,pst-text,pst-3d,multido,pst-plot,pst-coil}

\newtheorem{remark}{Remark}
\newtheorem{proposition}{Proposition}
\newtheorem{assumption}{Assumption}
\newtheorem{lemma}{Lemma}

\def\cala{{\cal A}}

\def\call{{\cal L}}
\def\lef[{\left[\begin{array}}
\def\rig]{\end{array}\right]}
\def\qed{\hfill$\Box \Box \Box$}

\def\rea{\mathbb{R}}

\usepackage{psfrag}

\newcommand{\col}{ \mbox{col} }
\newcommand{\rank}{ \mbox{rank } }

\def\cala{{\cal A}}

\def\call{{\cal L}}

\def\rea{\mathbb{R}}

\newcommand{\bfy}{\mbox{$y$}}

\def\begequ{\begin{equation}}
\def\endequ{\end{equation}}
\def\lab{\label}
\def\begite{\begin{itemize}}
\def\endite{\end{itemize}}
\def\begarr{\begin{array}}
\def\endarr{\end{array}}
\def\begequarr{\begin{eqnarray}}
\def\endequarr{\end{eqnarray}}

\usepackage{multicol}

\def\caly{{\cal Y}}

\def\calh{{\cal H}}

\def\cals{{\cal S}}

\def\calh{{\cal H}}

\def\call{{\cal L}}

\def\cala{{\cal A}}

\def\cald{{\cal D}}

\def\bfy{{\bf y}}

\def\bfp{{\cal P}}
\def\bfphi{\mathbb{\Phi}}

\def\bfphi{{\boldsymbol \phi}}

\def\hatthe{\hat{\eta}}
\def\tilthe{\tilde{\eta}}

\def\liminf{\lim_{t \to \infty}}

\def\intnum{\mathbb{Z}}

\def\hatthe{\hat{\theta}}
\def\tilthe{\tilde{\theta}}

\newcommand{\bfq}{\mbox{$q$}}

\def\cald{{\cal D}}

\def\hatthe{\hat{\eta}}
\def\tilthe{\tilde{\eta}}

\def\liminf{\lim_{t \to \infty}}

\def\L2{{\cal L}_2}
\def\L2e{{\cal L}_{2e}}

\def\rea{\mathbb{R}}

\def\reapos{\mathbb{R}_{>0}}
\def\reanonneg{\mathbb{R}_{\geq 0}}
\def\intnum{\mathbb{Z}}
\def\intnumpos{\mathbb{Z}_{>0}}
\def\intnumnonneg{\mathbb{Z}_{\geq 0}}
\def\sign{\mbox{sign}}
\def\adj{\mbox{adj}}

\def\hatthe{\hat{\theta}}
\def\tilthe{\tilde{\theta}}

\def\begmat#1{\begin{bmatrix}#1\end{bmatrix}}
\def\begali#1{\begin{align}{#1}\end{align}}
\def\begalis#1{\begin{align*}{#1}\end{align*}}

\def\begsubequ{\begin{subequations}}
\def\endsubequ{\end{subequations}}
\def\begequarr{\begin{eqnarray}}
\def\endequarr{\end{eqnarray}}
\def\begequarrs{\begin{eqnarray*}}
\def\endequarrs{\end{eqnarray*}}
\def\begarr{\begin{array}}
\def\endarr{\end{array}}
\def\begequ{\begin{equation}}
\def\endequ{\end{equation}}
\def\lab{\label}
\def\begdes{\begin{description}}
\def\enddes{\end{description}}
\def\begenu{\begin{enumerate}}
\def\begite{\begin{itemize}}
\def\endite{\end{itemize}}
\def\endenu{\end{enumerate}}

\def\lef[{\left[\begin{array}}
\def\rig]{\end{array}\right]}
\def\qed{\hfill$\Box \Box \Box$}
\def\begcen{\begin{center}}
\def\endcen{\end{center}}
\def\begrem{\begin{remark}\rm}
\def\endrem{\end{remark}}
\def\begassum{\begin{assumption}}
\def\endassum{\end{assumption}}
\def\begassums{\begin{assumption*}}
\def\endassums{\end{assumption*}}
\def\begassu{\begin{ass}}
\def\endassu{\end{ass}}
\def\beglem{\begin{lemma}}
\def\endlem{\end{lemma}}
\def\begcor{\begin{corollary}}
\def\endcor{\end{corollary}}
\def\begfac{\begin{fact}}
\def\endfac{\end{fact}}

\def\TAC{{\it IEEE Trans. Automatic Control}}

\def\IJC{{\it International Journal of Control}}
\def\AJC{{\it Asian Journal of Control}}

\def\SCL{{\it Systems and Control Letters}}
\def\AUT{{\it Automatica}}

\def\CSM{{\it IEEE Control Systems Magazine}}

\def\ARC{{\it IFAC Annual Reviews in Control}}
\def\CSL{{\it IEEE Control Systems Letters}}

\newtheorem{definition}{Definition}

\begin{document}
\title{Identifiability Implies Robust, Globally Exponentially Convergent On-line Parameter Estimation: Application to Model Reference Adaptive Control}
\author{Lei Wang, Romeo Ortega, Alexey Bobtsov, Jose Guadalupe Romero and Bowen Yi
\thanks{L. Wang and B. Yi are with the Australian Center for Field Robotics, The University of Sydney, NSW 2006, Sydney   (e-mail: lei.wang2$\{$bowen.yi$\}$@sydney.edu.au). }
\thanks{R. Ortega and J.G. Romero are with the  Departamento Acad\'emico de Sistemas Digitales, ITAM, R\'io Hondo 1, Ciudad de M\'exico, 01080, M\'exico (e-mail: romeo.ortega$\{$jose.romerovelaquez$\}$@itam.mx). }
\thanks{ A. Bobtsov  is with the Faculty of Control Systems and Robotics, ITMO University, Kronverkskiy av. 49, St. Petersburg, 197101, Russia  (e-mail: bobtsov@mail.itmo.ru).}
}
\maketitle
\begin{abstract}
In this paper we propose a new parameter estimator that ensures {\em global exponential convergence} of linear regression models requiring only the necessary assumption of {\em identifiability} of the regression equation, which we show is equivalent to interval excitation of the regressor vector. Continuous and discrete-time versions of the estimators are given. An extension to---separable and monotonic---{\em non-linear parameterizations} is also given.   The estimators are shown to be {\em robust} to additive measurement noise and---not necessarily slow---parameter variations. Moreover, a version of the continuous-time estimator that {\em rejects} sinusoidal disturbances with unknown internal model is given. The estimator is shown to be applicable to the classical model reference adaptive control problem {\em relaxing} the conspicuous assumption of known sign of the high-frequency gain. Simulation results that illustrate the performance of the estimator are given.
\end{abstract}


\section{Introduction}
\lab{sec1}
%
The tasks of control, identification or prediction of the dynamics of an unknown nonlinear system is usually accomplished assuming that there exists an approximation to the true dynamics with a fixed vector of parameters that globally fits the dynamics.  A typical scenario is to assume the dynamics is described by an ordinary differential (or difference) equation with unknown parameters, which are then estimated designing an on-line parameter estimator. In its simplest formulation it is assumed that these parameters enter {\em linearly} in the dynamic model leading to a relationship of the form $y=\phi^\top\theta$, with $y \in \rea,\;\phi\in \rea^q$ measurable signals and $\theta \in \rea^q$ a constant vector of unknown parameters---that we call in the sequel linear regression equation (LRE) with $\phi$ the regressor vector. LREs associated with many control problems, including system identification \cite{LJUbook}, adaptive control \cite{IOASUNbook,NARANNbook,SASBODbook}, filtering and prediction  \cite{GOOSINbook}, reinforcement learning  \cite{LEWetal} and sparse regression analysis \cite{BRUPROKUT}, have been reported in the literature.

Very often, gradient descent-based or least-squares parameter adaptation algorithms are used to generate on-line estimates of the unknown parameters. This leads to a linear time-varying (LTV) dynamical system that describes the behavior of the estimation errors, called parameter error equations (PEE), that have been extensively studied in the literature. A fundamental result is that a {\em necessary and sufficient} condition for the global {\em exponential} stability (GES) of the PEEs is that the regressor vector satisfies a { persistency of excitation} (PE) condition---which is a uniform complete observability property for the associated LTV system  \cite{IOASUNbook,NARANNbook,SASBODbook}. Here we underscore the qualifier ``exponential" because it is widely accepted that without this property parameter convergence cannot be ensured and the robustness of the schemes is seriously damaged. Moreover, without PE the transient performance of the estimators is highly unpredictable and only a weak monotonicity property of the estimation errors norm can be guaranteed.

This fragility situation is particularly clear in model reference adaptive control (MRAC), for which it has vividly been shown in \cite{IOAKOK,ROHetal} that several instability mechanisms are present.  In spite of intensive research efforts \cite{IOASUNbook,NARANNbook} the various fixes that have been introduced in the estimators---that include projections, deadzones and integrator leakages---have only partially alleviated this problem. Indeed, as shown in \cite{KHAORT}, it has only been possible to establish a ``continuity" property with respect to unmodeled dynamics and the preservation of signal boundedness in the face of noise. More precisely, it has been proven that there exists a sufficiently small bound on the norm of the error dynamics such that (some kind of) stability is preserved---but, unfortunately, this bound is not quantifiable from the data of the problem. Regarding the presence of noise or parameter variations only signal boundedness, again with no uniform bound, is established. It should be furthermore added that all of this ``robustified" schemes rely on the introduction of the dynamic normalization introduced in \cite{EGA}---and its importance for robustness established in \cite{ORTPRALAN,PRAyale}---that, as thoroughly discussed in \cite{MORijacsp},  slows-down the adaptation bringing some additional robustness problems.

Unfortunately, the PE property---which imposes a ``spanning behavior" to the signals---is rarely satisfied in applications, where the task is often to drive the signals to some constant value. Although it has recently been shown that (non-uniform) global asymptotic stability can be ensured under weaker assumptions \cite{BARORT,PRA}, these  conditions are still exremely stringent for applications. Hence the interest to propose new adaptation algorithms that ensure, via GES, parameter convergence without PE. This research line has been intensively pursued in the last few years and some recent adaptive schemes, where the PE assumption is obviated, have been reported in the literature---see  \cite{ORTNIKGER} for a recent survey.

To the best of the authors' knowledge the first estimators where parameter convergence is guaranteed under the extremely weak assumption of {\em interval excitation} (IE)  \cite{KRERIE}---called initial excitation in  \cite{PANetal} and excitation over a finite interval in \cite{TAObook}---are the concurrent  and the composite learning schemes reported in \cite{CHOetal} and \cite{PANYU}, respectively. These algorithms, which incorporate the monitoring of past data to build a stack of suitable regressor vectors, are closer in spirit to {\em off-line} estimators. See also  \cite{KRAKHA,ORTproieee} for two early references where a similar idea is explored. As is well-known, the main drawback of off-line estimators is their inability to track parameter variations, which is very often the main objective in applications. This situation motivates the interest to develop {\em bona-fide} on-line estimators that preserve the scheme's alertness.

Parameter convergence under the IE assumption was also recently established for the scheme reported in \cite{GERetalsysid}, that has the additional feature of ensuring  convergence in {\em finite-time}---see also \cite[Propositions 6 and 7]{ORTetaltac} and \cite{ORTBOBNIK}. A potential drawback of this algorithm is that it critically relies on the inclusion of a dynamic extension that mimicks the dynamics of the PEE, which may adversely affect the robustness of the estimator,  \cite[Remark 7]{ORTetaltac} and  \cite{ORTajc}. A similar difficulty is present in the algorithm recently proposed in \cite{WUetal}. \\

In this paper we are interested in the solution to the following key problems (KP).\\

\noindent {\bf KP1} Design an on-line estimator that ensures GES of the PEE under the weakest assumption that the LRE is {\em identifiable}.\\

\noindent {\bf KP2} Prove that, with a slight variation of the estimator that solves  {\bf KP1}, it is possible to prove the following important features:
\begite
\item  {\em Robustness}---in a clear, quantifiable sense---to external disturbances and (not necessarily slow) parameter variations.
\item {\em Rejection} of sinusoidal disturbances with {\em unknown} internal model.
\item Applicability to a well-defined class of {\em nonlinearly parameterized} regressor equations (NLPRE).
\endite

The first solution to {\bf KP1} was given in the recent paper  \cite{KORetal}  with the standing assumption that the regressor $\phi$ is IE, which we show in this paper is {\em equivalent} to identifiability of the LRE. Instrumental for the development of the adaptation algorithm proposed in \cite{KORetal} are the following steps. \\

\noindent {\bf S1} The use of the  {\em dynamic regressor extension and mixing} (DREM) parameter estimation procedure,  which was first proposed in \cite{ARAetaltac} for continuous-time (CT) and in \cite{BELetalsysid} for discrete-time (DT) systems. The construction of DREM estimators proceeds in two steps, first, the inclusion of a {\em free, linear operator} that creates an extended matrix LRE similar to the ones designed in \cite{KRE,LIO}. Second, a {\em nonlinear} manipulation of the data that allows to generate, out of an $q$-dimensional LRE,  $q$ {\em scalar}, and independent, LREs. DREM estimators have been successfully applied in a variety of identification and adaptive control problems, both, theoretical and practical ones, see \cite{ORTNIKGER,ORTetaltac} for an account of some of these results.\\

\noindent {\bf S2} The utilization of a procedure---proposed in \cite{BOBetal}---to generate, from a scalar LRE, new  scalar LREs where the {\em new regressor} satisfies some excitation conditions, even in the case when the original regressor is not exciting. To achieve this objective the authors borrow the key idea of the {\em generalized parameter estimation based observer} (GPEBO) \cite{ORTetalscl,ORTetalaut}, to generate the new LRE that includes some {\em free} signals. Then, applying the energy pumping-and-damping injection principle of \cite{YIetal}, these signals are selected  to guarantee some excitation properties of the new regressor. Unfortunately, to prove in \cite{BOBetal} that the aforementioned excitation properties guarantee GES it is necessary to assume some {\em a priori} non-verifiable  conditions   \cite[Proposition 3]{BOBetal}---in particular the absolute integrability of a signal and a non-standard requirement on the limiting behavior of some of the components of the trajectories of the estimator. Via the suitable selection of the aforementioned free signals in the new LRE, these two assumptions are relaxed in  \cite{KORetal} providing a definite answer to  {\bf KP1}. Recalling the procedure followed in the construction of the estimator of  \cite{KORetal}, that is, first the application of DREM and then invoke GPEBO, we refer to it in the sequel as D+G. Interestingly, for the new estimator we also rely on the use of GPEBO and DREM, but under different circumstances and used in the {\em opposite order}, hence we refer to it in the sequel as G+D. \\

In this paper we provide an answer to the more challenging {\bf KP2}, with our main contributions summarized as follows.\\

\noindent {\bf C1} We prove, for the first time, that IE of the original LRE is {\em equivalent} to identifiability of the parameters. That is, to the existence of $q$ linearly independent regressor vectors for the reconstruction of an $q$-dimensional parameter vector.\\

\noindent {\bf C2} The stability mechanisms and, consequently, the stability analysis of the G+D scheme is much more transparent than the ones of the D+G estimator. There are two consequences of this fact, on one hand, the procedure of {\em tuning} the estimator to achieve a satisfactory transient performance, which is difficult for the D+G scheme, is straightforward for the G+D one. On the other hand, by rendering the material accessible to a wider audience, the range of practical applicability of the new estimator is increased.\\

\noindent {\bf C3}  The {\em numerical complexity} of the proposed estimator is considerably simpler than the D+G scheme. In particular, GPEBO is applied to the PEE of the classical gradient estimator avoiding the reference to the, rather obscure, concept of ``virtual dynamics" used in the D+G estimator. Furthermore, the key mixing step of the DREM procedure reduces to a matrix multiplication, avoiding the need of generation of an extended LRE via the inclusion of additional LTV operators.\\

\noindent {\bf C4}  The estimators are shown to be {\em robust} to additive measurement noise and---not necessarily slow---parameter variations. This feature is established showing that the estimator may be derived applying the DREM technique, which is the action of a linear operator on the original LRE. Moreover, a variation of the CT estimator that {\em rejects} sinusoidal disturbances with unknown internal model is given. The qualifier ``reject" in the present context means that it is possible to have a {\em consistent estimate} of the unknown parameters $\theta$ in spite of the presence of the disturbances.\\

\noindent {\bf C5} The estimator is shown to be applicable to the classical MRAC problem, {\em relaxing} the conspicuous assumption of known sign of the high-frequency gain. As thoroughly discussed in \cite[Subsection 1.2]{ORTetal_aut19}---see also \cite[Section 3]{BARORTacsp18}---this key assumption is hard to verify in practice, and the schemes that avoid it are, either technically unsound \cite[Subsection 4.5.2]{IOASUNbook} or only of theoretical interest, since their transient performance is intrinsically bad and practically inadmissible \cite{NUS}.  \\

\noindent {\bf C6} Besides the case of LRE we consider  (separable and monotonic) {\em NLPRE}, with the associated estimator preserving all the properties of the case of LRE.  \\

\noindent {\bf C7} The behaviour of many physical systems is described via CT models. On the other hand, DT implementations of estimators are of significant {practical} relevance. Therefore, similarly to  \cite{KORetal,ORTetaltac}, to comply with both scenarios we consider in the paper both kinds of LREs. Interestingly, in contrast to  \cite{KORetal}, the construction and analysis tools of both cases are essentially the same.\\

The remainder of the paper is organized as follows.   In Section \ref{sec2} we prove the equivalence between IE of the regressor and identifiability of the parameters of the LRE. Section \ref{sec3} contains our main result for LRE. The proof that the proposed G+D estimator may be derived applying the DREM technique is given in Section \ref{sec4}. This important result is then used in Section \ref{sec5} to carry-out the robustness analysis, including the proof of BIBO-stability and disturbance rejection. In section \ref{sec6} we apply the G+D estimator to relax the key assumption of known sign of the high-frequency gain in MRAC. In Section \ref{sec7} we extend the results for a class of NLPRE.  Section \ref{sec8} presents some simulation illustrating our main results. The paper is wrapped-up with concluding remarks and future research in Section \ref{sec9}. To simplify the reading, some preliminary lemmata are given in the Appendix  and a list of acronyms is included at the end of the paper.\\

\noindent {\bf Notation.} $I_n$ is the $n \times n$ identity matrix and ${\bf 0}_{n\times q}$  is an $n \times q$ matrix of zeros. $\rea_{>0}$, $\rea_{\geq 0}$, $\intnum_{>0}$ and $\intnum_{\geq 0}$ denote the positive and non-negative real and integer numbers, respectively. For $x \in \rea^n$, we denote saure of the Euclidean norm as $|x|^2:=x^\top x$.  Given $q \in \intnum_{>0}$ we define the set $\bar q:=\{1,2,\dots,q\}$. CT signals $s:\rea_{\geq 0} \to \rea$ are denoted $s(t)$, while for DT sequences $s:\intnum_{\geq 0} \to \rea$ we use $s(k)$.  When a formula is applicable to CT signals and DT sequences the time argument is {\em omitted}. The symbol $\|\cdot\|_\infty$ stands for the infinity norm of a signal or sequence. The action of an operator $\mathcal H$ on a CT signal $s(t)$ is denoted as $\mathcal H[s](t)$, and $\mathcal H[s](k)$ for a sequence $s(k)$. In particular, we define the derivative operator $\bfp^n[s](t)=:{d^n s(t)\over dt^n}$ and the delay operator  $\bfq^{\pm n}[s](k)=:s(k \pm n)$, where $n \in \intnum_{ >0}$.
%
\section{Interval Excitation is Equivalent to Identifiability}
\lab{sec2}
%
Throughout the paper we deal with LRE of the form
\begequ
\lab{lre}
y=\phi^\top \theta,
\endequ
where $y \in \rea,\;\phi\in \rea^q$ are {\em measurable} signals and $\theta \in \rea^q$ is a constant vector of {\em unknown} parameters.\footnote{To simplify the notation we consider the case of {\em scalar} $y$, as will become clear below, the extension to the matrix case is straightforward.}  The main objective of the paper is to provide a solution to {\bf KP2}. To streamline the main result we need the following definition.

\begin{definition}
\lab{def1}
A bounded signal $\phi \in \rea^{q}$ is IE \cite{KRERIE,TAObook} if
\begalis{
	&\int_0^{t_c} \phi(s) \phi^\top(s)  ds \ge C_c I_q,	\\
}
for some  $C_c \in \rea_{>0}$ and $t_c \in \rea_{>0}$ in CT and
\begalis{
	&\sum_{j=0}^{k_d} \phi(j) \phi^\top(j) \ge C_d I_q,
}
for some  $C_d \in \rea_{>0}$ and $k_d \in \intnum_{> 0}$ in DT.
\end{definition}

In this section we prove the fundamental result that IE of $\phi$ is equivalent to identifiability of the LRE  \eqref{lre}. We recall that identifiability, which is defined below, is a {\em necessary and sufficient} condition to reconstruct (even off-line) the unknown parameters.

\begin{definition}
\lab{def2}
The LRE \eqref{lre} is said to be {\em identifiable} if and only if there exists a set of time instants---$\{t_i\}_{i\in \bar q},\;t_i \in \reapos$ in CT and $\{k_i\}_{i\in \bar q},\;k_i \in \intnumpos$ in DT---such that
$$
\rank\Big\{\begmat{\phi(\tau_1)|\phi(\tau_2)|&\cdots&|\phi(\tau_q)}\Big\}=q,
$$
where $\tau_i=t_i$ in CT and $\tau_i=k_i$ in DT.
\end{definition}

\begin{proposition}
\lab{pro1}
The LRE \eqref{lre} is identifiable {\em if and only if} the regressor vector $\phi$ is IE.
\end{proposition}
\begin{proof}
The proof of the DT version is obvious recalling that for any symmetric matrix $A \in \rea^{q \times q}$ we have the following equivalence
$$
A>0 \;\Leftrightarrow\; z^\top  A z>0,\;\forall z \in \rea^q \setminus \{0\}
$$
that, given the definition of IE, imposes the constraint $k_d \geq q$.

The proof of the CT case proceeds as follows. The necessity is proved by contradiction. We suppose that there exists a positive integer $q_0 <q$ such that
\begequ
\lab{ranmat}
\mbox{rank}\big[\phi(t_1)|\ldots|\,\phi(t_q)\big] \leq q_0
\endequ
holds for all time sequence $\{t_i\}_{i\in \bar q}$, with $\{\bar t_i\}_{i\in \bar q}$ being such that
\[
\mbox{rank}\big[\phi(\bar t_1)|\ldots|\,\phi(\bar t_q)\big] = q_0\,.
\]
Let $h\in\mathbb{R}^q$ be such that $|h|=1$ and
\[
\phi^\top(\bar t_i)h=0,\; \forall {i\in \bar q}.
\]
Next we show that $\phi^\top(t)h=0$ for all $t \in \rea_{\geq 0}$ by contradiction. We suppose that there exists a $t^\sharp  \in \rea_{\geq 0}$ such that
\[
\phi^\top(t^\sharp)h\neq 0\,.
\]
This indicates
\[
q_0 = \mbox{rank}\big[\phi(\bar t_1)|\ldots |\,\phi(\bar t_q)\big] < \mbox{rank}\big[\phi(\bar t_1)|\ldots |\,\phi(\bar t_q)|\,\phi(t^\sharp)\big]
\]
which contradicts with the assumption that \eqref{ranmat} holds for all time sequence $\{t_i\}_{i\in \bar q}$. Hence, we have $\phi^\top(t)h=0$ for all $t  \in \rea_{\geq 0}$.

With this in mind, it can be easily seen that
\[
h^\top \int_{0}^{t_c}\phi(\tau)\phi^\top(\tau)d\tau h = \int_{0}^{t_c}|\phi^\top(\tau)h|^2d\tau = 0\,.
\]
This clearly contradicts with the IE condition. Therefore, there exists a time sequence $\{t_i\}_{i\in \bar q}$ such that $\mbox{rank}\big[\phi(t_1)|\ldots |\,\phi(t_q)\big] = q$, completing the necessity proof.

To prove sufficiency we let $t_c > t_q$, and proceed to show that
\[
\int_{0}^{t_c}\phi(\tau)\phi^\top(\tau)d\tau >0\,.
\]
As the matrix $\big[\phi(t_1)|\ldots |\,\phi(t_q)\big]$ is full rank with identifiability, it can be seen that for any $h\in\mathbb{R}^q$ satisfying $|h|=1$, there always exists a $\bar i\in\bar q$ such that
\[
|\phi(t_{\bar i})h| >0\,.
\]
By continuity, it follows that for any $h\in\mathbb{R}^q$ satisfying $|h|=1$, there exists  an $\epsilon>0$ such that
\[
\sum_{i\in\bar q}|\phi(\hat t_{i})h|\geq |\phi(\hat t_{\bar i})h|>0\,,\; \forall \hat t_{i}\in[t_i,t_i+\epsilon]\,
\]
yielding
\[
\int_{0}^{t_c}|\phi^\top(\tau)h|^2d\tau \geq \int_{t_{\bar i}}^{t_{\bar i}+\epsilon}|\phi^\top(\tau)h|^2d\tau >0\,.
\]
Therefore, by recalling that such $h$ is arbitrary, it can be concluded that  $\int_{0}^{t_c}\phi(\tau)\phi^\top(\tau)d\tau >0$ for $t_c > t_q$.
The proof is thus completed.
\end{proof}

\begrem
\lab{rem1}
For the sake of simplicity, we present $y$ and $\phi$ in \eqref{lre} as functions of time, in the understanding that they may be functions of measurable signals evaluated at time $t$ in CT or $k$ in DT, for instance, the state, input and/or output of a dynamical system---see \cite{ORTNIKGER} and Sections \ref{sec6} and \ref{sec8} for particular examples. Also, following standard practice in identification and adaptive control, in the sequel we disregard the presence of the exponentially decaying term stemming from the effect of the initial conditions of various filters used to generate the regression, see  \cite[Lemma 2]{ARAetalpe} where the effect of this term in the DREM estimator is rigorously analyzed.
\endrem
%
\section{Main Result for Linear Regression Equations}
\lab{sec3}
%
In this section we present the G+D estimators that solve {\bf KP2} in CT and DT for the LRE \eqref{lre}.

\begin{proposition}
\lab{pro2}
Consider the LRE \eqref{lre}. Define the {\em G+D interlaced} estimator
\begsubequ
\lab{intest}
\begali{
\lab{theg}
\mathfrak{H}_a[\hat \theta_g] & = \cala \hat\theta_g + \mathfrak{g} \phi y,\; \hat\theta_g(0)=\theta_{g0} \in \rea^q\\
\lab{phi}
\mathfrak{H}_b[\Phi]& =\cala\Phi,\;\Phi(0)=I_q\\
\lab{the}
\mathfrak{H}_a[\hat \theta] & =\mathfrak{d}\Delta(Y-\Delta \hat\theta),\; \hat\theta(0)=\theta_0 \in \rea^q,
}
\endsubequ
where the operators $\mathfrak{H}_a[\cdot]$ and  $\mathfrak{H}_b[\cdot]$ are defined as
\begequ
\lab{oped}
\mathfrak{H}_a[s]:=\left\{ \begarr{ccl} \bfp[s](t) & \mbox{in} & CT \\ &&\\ (\bfq-1)[s](k) & \mbox{in} & DT \endarr \right.,\;
\endequ
\begequ
\mathfrak{H}_b[s]:=\left\{ \begarr{ccl} \bfp[s](t) & \mbox{in} & CT \\ &&\\ \bfq[s](k) & \mbox{in} & DT \endarr \right.,\;
\endequ
the functions $\mathfrak{g},\mathfrak{d}$ and $\cala$ are given by
$$
\mathfrak{g}:=\left\{ \begarr{ccl} \gamma_g(t) & \mbox{in} & CT \\ &&\\ {1 \over \gamma_g(k)+|\phi(k)|^2} & \mbox{in} & DT \endarr \right.,\;
\mathfrak{d}:=\left\{ \begarr{ccl} \gamma & \mbox{in} & CT \\ &&\\  {1 \over \gamma+\Delta^2(k)} & \mbox{in} & DT \endarr \right.
$$
\begequ
\lab{cala}
\cala :=\left\{ \begarr{ccl}-\mathfrak{g}(t) \phi(t) \phi^\top (t) & \mbox{in} & CT \\ &&\\ I_q-\mathfrak{g}(k)\phi(k)\phi^\top (k) & \mbox{in} & DT \endarr \right.\\
\endequ
with $\gamma_g(\cdot)  \in \rea_{> 0}$, $\gamma  \in \rea_{> 0}$,  and the functions
\begsubequ
\lab{aydel}
\begali{
\lab{cald}
   \cald &:=I_q-\Phi \\
\lab{del}
\Delta & :=\det\{\cald\}\\
\lab{y}
Y & := \adj\{\cald\} (\hat\theta_g-\Phi\theta_{g0}),
}
\endsubequ
where $\adj\{\cdot\}$ stands for the adjugate matrix. If the LRE \eqref{lre} is {\em identifiable} then
$$
\lim_{\tau \to \infty}\tilthe(\tau)=0,\;(exp)
$$
where $\tilthe:=\hatthe-\theta$ and $\tau=t$ in CT and $\tau=k$ in DT.
\end{proposition}

 \begin{proof}
Replacing \eqref{lre} in \eqref{theg} yields the PEE for the gradient estimator
$$
\mathfrak{H}_a[{\tilde \theta}_g]  =\cala \tilde\theta_g,
$$
where $\tilthe_g:=\hatthe_g-\theta$, and we used the definition   \eqref{cala}.   Consequently, from  the properties of the {\em fundamental matrix}  $\Phi$ \cite{RUGbook} defined in \eqref{phi}, we get
\begequ
\lab{dyntiltheg}
	 \tilde \theta_g  =\Phi \tilde\theta_g(0),
\endequ
which may be rewritten as the extended LRE
\begali{
\lab{keyide}
	  \cald\theta &=\hat \theta_g -\Phi \theta_{g0}\,,
}
where we used \eqref{cald}. Following the DREM procedure we multiply \eqref{keyide} by   $\adj\{\cald\}$ to get the following {\em scalar} LRE
\begequ
\lab{ydel}
Y_i = \Delta\theta_i,\;\quad i \in \bar q,
\endequ
where we used \eqref{del} and \eqref{y}. We underscore the fact that the regressor $\Delta$ is a {\em scalar}.

Replacing \eqref{ydel} in \eqref{the} yields the PEE for each of the elements $\tilde\theta_i,\;i \in\bar q$, of the vector  $\tilde\theta$ of the  least mean squares estimator \eqref{the}
\begali{
\lab{peethe}
\mathfrak{H}_a[\tilde \theta_i] & =-\mathfrak{d} \Delta^2\tilde\theta_i.
}
Now, in Proposition \ref{pro1} it is shown that identifiability is equivalent to $\phi$ in IE. On the other hand, in Lemmas \ref{lem3} and \ref{lem5}, given in Appendix A, we prove that the IE assumption implies that  $\Delta$ is PE in the CT and DT case, respectively. The proof  of exponential  convergence in CT follows from  the well-known result \cite[Theorem 2.5.1]{SASBODbook}.

For the DT case we have the following argument. The PEEs for the normalized least mean squares estimator \eqref{the} are given by
\begalis{
\tilde \theta_i(k+1) & =\tilde \theta_i(k)-{ \Delta^2(k) \over \gamma+ \Delta^2(k)}\tilde\theta_i(k)\\
 & ={ \gamma \over \gamma+ \Delta^2(k)}\tilde\theta_i(k)\\
 & = \prod^k_{j=0}{ \gamma \over \gamma+ \Delta^2(j)}\tilde\theta_i(0).
}
From the fact that $\Delta(k)\in \rea_{>0}$ for all $k\geq k_d$ we conclude that  $\Delta(k) \notin \ell_2$. This, together with the fact that
$$
\Delta(k) \notin \ell_2 \; \Leftrightarrow\;  \prod^\infty_{j=0}{ \gamma \over \gamma+ \Delta^2(j)}=0,
$$
proves global convergence. The proof that the  convergence is exponential follows from the inequality
$$
 \prod^k_{j=0}{ \gamma \over \gamma+ \Delta^2(j)} \leq \exp\Big(- \sum^k_{j=0}{  \Delta^2(j) \over \gamma+ \Delta^2(j)} \Big).
 $$
\end{proof}

\begrem
\lab{rem2}
It is important to underscore that the estimator of Proposition \ref{pro1} consists of the interlacing of two classical gradient-based parameter search algorithms and contains only two tuning gains $\gamma$ and $\gamma_g$. The effect of both gains on the transient performance of the estimator is very clear and has been extensively studied in the literature---see \cite{EFIFRA} for a recent survey of the main results on this topic. The importance of this fact can hardly be underestimated because, as is well-known, the stage of commissioning the estimators, which is usually done with a trial-and-error approach, is very painful and a bad tuning has a serious deleterious effect on the overall performance of the scheme.
\endrem

\begrem
\lab{rem3}
Notice that in the interval $t\in[0,t_c)$ in CT or $k \in [0,k_d)$ in DT the IE condition is not yet satisfied, which implies that $\Delta=0$ in this interval. Consequently, the second estimator remains {\em frozen} in this interval, that is ${\hat\theta}={\theta}_0$. We can, therefore, interpret the role of both estimators as follows: the one of $\hat \theta_g$ ``gathers" the required excitation while the one of $\hat \theta$ starts ``operating" only after we have a rich regressor.
\endrem
%
\section{A DREM Perspective of the Proposed Estimators}
\label{sec4}
%
In this section we show that the reparameterization \eqref{keyide} used in the estimator of Proposition \ref{pro2} can be generated applying the well-known DREM procedure  \cite{ARAetaltac} to the LRE \eqref{lre} in both, the CT and the DT cases. Towards this end, we recall that the first step in DREM is to generate an extended regressor applying a {\em linear, single-input $q$-output} operator $\calh$ to the LRE \eqref{lre}. Because of linearity, this yields the new (extended) LRE
\begin{equation}\label{eq:new-LRE}
\bfy= \bfphi \theta
\end{equation}
where we defined the vector $\bfy \in \rea^q$ and the matrix $\bfphi \in \rea^{q \times q}$ as
\begalis{
\bfy  &:= \calh[y] \\
\bfphi & :=\begmat{\calh[\phi_1]|  \calh[\phi_2] |\cdots  |\calh[\phi_q]}.
}
In the proposition below we identify an LTV operator $\calh$ such that
\begalis{
\bfy &=\hat \theta_g -\Phi \theta_{g0}\\
\bfphi& =\cald,
}
yielding the extended LRE \eqref{keyide}, hence proving the claim above.
\begin{proposition}
\lab{pro3}
Define the single-input $q$-output operator $\calh:u \mapsto {\bf u}$, with $u \in \rea$, ${\bf u} \in \rea^q$, and state-space representation
\begali{
\nonumber
\mathfrak{H}_a[x_u] & =\cala  x_u  + \mathfrak{g} \phi u \\
\lab{opeh}
{\bf u} &=x_u,
}
where the operator $\mathfrak{H}_a$ and the functions $\cala$ and $\mathfrak{g}$ are defined in Proposition \ref{pro2} and the initial condition of the state $x_u$ is zero. Applying this operator to the signal of the LRE \eqref{lre} we obtain the LTV systems
\begin{equation}\label{eq:H-op}\begin{array}{rcl}
\mathfrak{H}_a[x_y] & =&\cala  x_y  + \mathfrak{g} \phi  y \\
\mathfrak{H}_a[x_{\phi,i}] &=&\cala  x_{\phi,i}   + \mathfrak{g}\phi  \phi_i,\;\quad i \in\bar q,
\end{array}\end{equation}
with initial conditions {$x_y(0)={\bf0}_{q \times 1}$ and $x_{\phi,i}(0)={\bf0}_{q \times 1},\;i \in\bar q$}. Then, for all $t  \in \reanonneg$ in CT and all $k \in \intnumnonneg$ in DT, the following identity holds
\begali{
\nonumber
\hat \theta_g {-\Phi \theta_{g0} } &= x_y \\
\lab{tilsig}
\cald&={\begmat{x_{\phi,1} \,|\,  x_{\phi,2} \,|\,\cdots  \,|\,x_{\phi,q}}}\,.
}
\end{proposition}
 \begin{proof}
First, notice that  we have
\begalis{
\mathfrak{H}_b[\cald] &= \mathfrak{H}_b[I_q]-\mathfrak{H}_b[\Phi] \\
 & =  \mathfrak{H}_b[I_q]- \cala\Phi\\
  & =  \mathfrak{H}_b[I_q] - \cala[I_q- \cald]\\
    & =    \cala \cald- \cala + \mathfrak{H}_b[I_q]\\
      & =  \cala \cald+ \mathfrak{g}\phi\phi^\top
}
where we have used \eqref{cala} and the fact that $\mathfrak{H}_b[I_q]=0$ in CT and $\mathfrak{H}_b[I_q]=I_q$ in DT to derive the last equality.

Now, define the state vector errors
\begalis{
\tilde x_y &:= x_y - (\hat \theta_g-\Phi\theta_{g0})\\
\tilde x_{\phi,i} &:=x_{\phi,i} - \cald_i,\;i \in\bar q
}
with $\cald_i$ the $i$-th column of $\cald$ and notice that
\begalis{
\mathfrak{H}_a[\tilde x_y] &= \cala \tilde x_y \\
\mathfrak{H}_a[\tilde x_{\phi,i}] &=\cala \tilde x_{\phi,i},\;i \in\bar q.
}
The proof is completed noting that  $\tilde x_y(0)={\bf0}_{q \times 1}$ and $\tilde x_{\phi,i}(0)={\bf0}_{q \times 1}$, hence \eqref{tilsig} holds true for all $t  \in \reanonneg$ in CT and all $k \in \intnumnonneg$ in DT.
  \end{proof}

\begin{remark}
\lab{rem4}
In DREM it is usually assumed that the operator $\calh$ is bounded-input bounded-output (BIBO)-stable. This condition is imposed to preserve boundedness of the extended LRE \eqref{eq:new-LRE}, which however is not necessary for the overall stability analysis of the DREM estimator. It is possible to show that---without further assumptions on $\gamma_g$, besides positivity---the operator  $\calh$, defined via \eqref{opeh}, is {\em not BIBO-stable}.
\end{remark}
%
\section{Robustifying the Proposed Estimators}
\label{sec5}
%
In this section we analyze the robustness {\em vis-\`a-vis} additive perturbation of (a slight variation of) the estimator of Proposition \ref{pro2}. That is, we consider the {\em perturbed} LRE
\begequ
\lab{perlre}
  y=   \phi^\top  \theta + d,
\endequ
where $d$ represents an additive perturbation signal. This signal may come from additive noise in the measurements of $y$ and $\phi$ or time variations of the parameters, that is, $d$ may be decomposed as
$$
d=d_y + d^\top_\phi \theta + d^\top_\theta \phi,
$$
where $d_y\in \rea$ and  $d_\phi \in \rea^q$ represent the measurement noise added to $y$ and $\phi$, respectively, and $d_\theta \in \rea^q$ captures {\em time variations} in the parameters. We make the reasonable assumption that these signals are all bounded.
\subsection{BIBO stability}
\lab{subsec51}
%
In this subsection we prove that, imposing an {\em additional condition} on the adaptation gain $\gamma_g$, it is possible to robustify the proposed estimators with respect to  the additive disturbance $d$. More precisely, we will prove the parameter estimation error $\tilde \theta$ remains bounded. Instrumental to establish these results are, on one hand the claim of GES of the unperturbed estimator of Proposition \ref{pro2} and, on the other hand, the proof in Proposition \ref{pro3}  that the estimators can be derived from the DREM procedure. As indicated in Remark \ref{rem3} the operator $\calh$---defined with {\em arbitrary, positive} $\gamma_g$ in \eqref{opeh}---is not BIBO-stable, hence in this subsection an additional constraint on $\gamma_g$ is imposed to ensure that the operator $\calh$ is BIBO-stable. Consequently, the presence of the disturbance $d$ will induce an additive {\em bounded} disturbance on the extended LRE \eqref{keyide}, yielding a GES system with a bounded perturbation.

The main result is summarized in the proposition below.

\begin{proposition}
\lab{pro4}
Consider  the {\em perturbed} LRE \eqref{perlre} with bounded $d$. Assume the unperturbed LRE \eqref{lre} is identifiable. If the adaptation gain $\gamma_g  \in \rea_{> 0}$ is selected such that $\int_{0}^{\infty}\gamma_g(t)dt$ is bounded in CT or $\sum_{k=0}^{\infty}\frac{1}{\gamma_g(k)}$  is bounded in DT, the G+D estimator of Proposition \ref{pro2} is robust in the sense that the parameter estimation error $\tilde \theta$ remains bounded.
\end{proposition}

\begin{proof}
Applying the operator $\calh$ of Propositions \ref{pro3}  to the perturbed LRE yields the perturbed version of the extended LRE \eqref{keyide} as
\begin{equation}\label{eq:per-LRE}
\hat \theta_g -\Phi \theta_{g0}= \cald \theta + \calh[d],
\end{equation}
where we exploited the property of linearity of $\calh$. Next we proceed to show that, under the conditions on $\gamma_g$ imposed in the proposition, the operator $\calh$ is BIBO-stable. This is done by proving that, for all bounded $d$, the signal $\calh[d]$ is also bounded.

For the CT case, the signal $\calh[d](t)$ is generated via the CT LTV system
\begalis{
\dot x_d(t)& = -\gamma_g(t) \phi(t) \phi^\top (t) x_d(t) + \gamma_g(t)\phi(t)d(t)\\
\calh[d](t)&=x_d(t).
}
Defining $V(x_d) := |x_d|^2$, we have
\begalis{
\dot V & \leq -2\gamma_g(t) |\phi^\top (t) x_d(t)|^2 + 2\gamma_g(t)|\phi^\top (t) x_d(t)||d(t) | \\
& \leq -\gamma_g(t) |\phi^\top (t) x_d(t)|^2+\gamma_g(t)|d(t) |^2,
}
which yields
\begalis{
V(t) - V(0)& \leq \int_{0}^{t}\gamma_g(\tau)|d(\tau) |^2 d\tau\\
& \leq \|d(t)\|^2_\infty\int_{0}^{t}\gamma_g(\tau)d\tau.
}
This further implies that  $x_d(t)$, hence $\calh[d](t)$, is bounded as $d(t)$  and $\int_{0}^{\infty}\gamma_g(t)dt$ are bounded.\\

For the DT case, the signal $\calh[d](k)$ is generated via the DT LTV system
\[
x_d(k+1) = \left(I_q-\frac{\phi(k) \phi^\top (k)}{\gamma_g(k)+|\phi(k)|^2}\right) x_d(k) + \frac{\phi(k) d(k)}{\gamma_g(k)+|\phi(k)|^2}.
\]
Similarly to the CT case, for the function $V(x_d)$ we  obtain
\[
V(k+1) - V(k) \leq 4\frac{|d(k)|^2}{\gamma_g(k)}
\]
which yields
\begalis{
V(k+1) - V(0) & \leq \sum_{j=0}^{k}\frac{4|d(j)|^2}{\gamma_g(j)}\\
& \leq \|d(k)\|_\infty^2\sum_{j=0}^{k}\frac{4}{\gamma_g(j)}.
}
As $d(k)$ and $\sum_{k=0}^{\infty}\frac{1}{\gamma_g(k)}$ are bounded this implies that $x_d(k)$ is bounded.

From the analysis above, we conclude that the operator $\calh$ is BIBO-stable and $\calh[d]$ is bounded. Then, multiplying \eqref{eq:per-LRE} by   $\adj\{\cald\}$ we get the following perturbed LRE
\begequ
\lab{ydelper}
Y = \Delta\theta+\xi,
\endequ
where we defined the signal
\begequ
\lab{pertilthe}
\xi:= \adj\{\cald\}\calh[d].
\endequ
We notice that, due to Proposition \ref{pro3} and the BIBO-stability of $\calh$, this signal is bounded. Replacing \eqref{pertilthe} in the estimator \eqref{the} yields the scalar, perturbed PEEs
$$
\dot{\tilde \theta}_i(t)=-\gamma \Delta^2(t)\tilde\theta_i(t)+\gamma \Delta(t) \xi_i(t),\;i \in \bar q,
$$
in CT and
$$
\tilde \theta_i(k+1)  ={ \gamma \over \gamma+ \Delta^2(k)}\tilde\theta_i(k)+{ \gamma \Delta^2(k)\over \gamma+ \Delta^2(k)} \xi_i(k),\;i \in \bar q
$$
in DT. Notice that in both cases $\Delta$ is PE by the Lemmata \ref{lem3} and \ref{lem5} in the Appendix and thus we are dealing with GES scalar systems with bounded additive perturbations.  The proof of the claims follows then invoking standard arguments of GES systems with bounded additive perturbations \cite{JIAWAN,KELDOW}.
\end{proof}

\begin{remark}
\lab{rem5}
As indicated above the additional assumption on $\gamma_g$ is introduced to robustify the estimator of $\hat\theta_g$ in \eqref{theg} in the sense that $\hat\theta_g$ is bounded if the perturbation $d$ is bounded. We remark that this objective may be achieved with other robustified estimators methods, such as the dead-zone \cite{ORTLOZ} and projection \cite{SASBODbook,TAObook}.
\end{remark}

\begin{remark}
\lab{rem6}
It is important to remark that the boundedness constraint imposed on the adaptation gain $\gamma_g$ essentially imposes that the estimator of $\hat \theta_g$ {\em looses its alertnesss} properties. However, since we are dealing with an interlaced estimator that incorporates a second DREM-based stage, the overall scheme does not necessarily looses the alertness. In this respect, it would be more interesting to use, as suggested in Remark \ref{rem5} above, other robustification methods that do not suffer from this drawback.
\end{remark}
\subsection{Rejection of sinuoidal disturbances with unknown internal model for CT LRE}
\lab{subsec52}
%
In this subsection we consider the CT perturbed LRE \eqref{ydelper} under the assumption that the disturbances $\xi_i(t),\; i \in \bar q$, are  sinusoidal signals of the form
$$
\xi_i(t)= a_{i} \sin(\omega_{i} t + \psi_{i}),\; i \in \bar q,
$$
with unknown amplitudes  $a_{i}\in \rea_{>0}$, frequencies $\omega_{i}\in \rea_{>0}$ and phase shifts $\psi_{i}\in \rea$.

Recalling that $\Delta(t)$ is a scalar signal, \eqref{ydelper} consists of a set of scalar perturbed LREs of the form
\begequ
\lab{disLRE}
Y_i(t) = \Delta(t)\theta_i+\xi_i(t).
\endequ
The main result is the proof that, with a suitable dynamic extension and a second application of the GPEBO approach to each scalar entry, it is possible to derive an {\em unperturbed LRE} for $\theta_i$ to which we can apply the standard gradient or DREM estimators to reconstruct $\theta_i$.

The first step in the design is to, fix a constant $\lambda \in \rea_{>0}$, and apply the filter
$$
F(\bfp):={\lambda^2 \over (\bfp + \lambda)^2},
$$
to {\em each element} of \eqref{ydelper}, yielding the scalar, perturbed LRE\footnote{To simplify the notation, we omit throughout the subsection the subindex $i$.}
\begequ
\lab{filperlre}
Y_{\tt F}(t) = \Delta_{\tt F}(t)\theta+\xi_{\tt F}(t),
\endequ
where we defined the filtered signals $(\cdot)_{\tt F}:=F(\bfp)(\cdot)$. Notice that since
$$
\xi(t)=a\sin(\omega t + \psi),
$$
we have that $\ddot \xi_{\tt F}(t) = - \omega^2 \xi_{\tt F}(t)$, from which we obtain the parameterization
\begali{
\nonumber
\xi_{\tt F}(t) &=- {1 \over \omega^2}\ddot \xi_{\tt F}(t)\\
\nonumber
& =  {\theta \over \omega^2}\ddot \Delta_{\tt F}(t)- {1 \over \omega^2}\ddot Y_{\tt F}(t)\\
& =:  \varphi^\top(t)\mu,
\lab{ddotxif}
}
where we used \eqref{filperlre} to get the second identity, the unknown parameter vector $\mu$ is given as
\begequ
\lab{newpar}
\mu:=\col\Big({\theta\over \omega^2},{1\over \omega^2}\Big),
\endequ
 and we defined the {\em measurable} regressor
\begequ
\lab{varphi}
\varphi(t):=\col(\ddot \Delta_{\tt F}(t),-\ddot Y_{\tt F}(t)).
\endequ
Obviously, since $\xi_{\tt F} $ is unmeasurable, we cannot use \eqref{ddotxif} for the estimation of $\mu$. It is, at this point, where we use again GPEBO to overcome this difficulty---as shown in the proposition below.

\begin{proposition}
\lab{pro5}
Consider the scalar, perturbed LRE \eqref{filperlre}. Define the dynamic extension
\begsubequ
\lab{disrej}
\begali{
\lab{disrejz}
\dot z(t) &= -z(t) - Y_{\tt F}(t),\;z(0)=0\\
\lab{disrejr}
\dot r(t) & = \cala_\xi(t) r(t)  + b_\xi(t),\; r(0)=\col(0,0)\\
\lab{disrejome}
\dot \Omega(t) & =\cala_\xi(t) \Omega(t)-e_2 \varphi^\top(t),\;\Omega(0)={\bf 0}_{2\times 2}\\
\lab{disrejphi}
\dot \Phi_\xi(t)& =\cala_\xi(t) \Phi_\xi(t),\; \Phi_\xi(0)=I_2,
}
\endsubequ
where $\varphi(t)$ is given in \eqref{varphi}, and we defined the matrices
\begsubequ
\lab{disrejmat}
\begali{
\lab{disreja}
   \cala_\xi(t) &:=\begmat{0 & \Delta_{\tt F}(t) \\ -\Delta_{\tt F}(t) & -1} \\
\lab{disrejb}
 b_\xi(t) &:=\begmat{-\Delta_{\tt F}(t) z(t)\\ 0},
}
\endsubequ
and $e_2:=\col(0,1)$. The following LRE holds
\begequ
\lab{disrejlre}
z(t) - r_2(t) =\begmat{(\Phi_\xi)_{2,1}(t) & \Omega_{2,1}(t)  & \Omega_{2,2}(t)}\begmat{\theta \\ \mu_1 \\ \mu_2},
\endequ
where the unknown vector $\mu$ is defined in \eqref{newpar}.
\end{proposition}

\begin{proof}
Similarly to \cite{BOBetal} notice that, since $\theta$ is {\em constant}, we can write
$$
\dot \theta(t)=\Delta_{\tt F}(t) z(t) - \Delta_{\tt F}(t) z(t),
$$
while the $z$-dynamics \eqref{disrejz} may be written as
$$
\dot z(t) = -z(t) -\Delta_{\tt F}(t) \theta - \xi_{\tt F}(t),
$$
where we used \eqref{filperlre}. Combining these two equations define the ``virtual" dynamical system
\begalis{
\dot \chi(t) & =\cala_{\xi}(t) \chi(t)  - \begmat{\Delta_{\tt F}(t) z(t) \\ \xi_{\tt F}(t)}\\
  & =\cala_{\xi}(t) \chi(t) + b_\xi(t) - e_2 \varphi^\top(t) \mu
}
where the ``state" is $\chi(t):=\col(\theta(t),z(t))$ and we used \eqref{disreja} in the first identity and \eqref{ddotxif}, \eqref{varphi} and \eqref{disrejb} to get the second equation.

Define the signal
$$
e_\xi(t):=r(t)-\chi(t)+\Omega(t) \mu,
$$
which satisfies $\dot e_\xi(t)=\cala_\xi(t) e_\xi(t)$. Consequently, in view of the definition of $\Phi_\xi(t)$, we have that
\begalis{
e_\xi(t) & = \Phi_\xi(t) e_\xi(0)\\
& = \Phi_\xi(t) \begmat{-\theta \\ 0},
}
where we took into account the choice of initial conditions in \eqref{disrej}. From the equation above we have that $e_{\xi,2}(t)=-(\Phi_\xi)_{2,1}(t) \theta$. Now, taking the second element from the definition of $e_\xi(t)$ we get
\begalis{
z(t) &=r_2(t)- e_{\xi,2}(t) +\Omega_{2,1}(t)\mu_1 + \Omega_{2,2}(t) \mu_2\\
&= r_2(t) + (\Phi_\xi)_{2,1}(t) \theta +\Omega_{2,1}(t) \mu_1 + \Omega_{2,2}(t) \mu_2.
}
This completes the proof.
\end{proof}

\begrem
\lab{rem70}
It is important to note that the standing assumption in this subsection is that it is the LRE \eqref{ydelper}, which is the LRE generated by the G+D procedure, that is perturbed by  a sinusoidal disturbance. It is not clear which kind of disturbances $d(t)$ to the original LRE \eqref{perlre} will give rise to a sinusoidal signal $\xi(t)$ since the relation between these two signal involves complicated operations, namely  \eqref{pertilthe}.
\endrem

\begrem
\lab{rem7}
Replacing \eqref{newpar} in \eqref{disrejlre} we have that
$$
\col(\theta, \mu_1, \mu_2)=\col\Big(\theta, {\theta\over \omega^2},{1\over \omega^2}\Big),
$$
hence there are only two unknown parameters---$\theta$ and  $\omega$---and the LRE is {\em overparameterized}. In Section \ref{sec7} we give a procedure to identify the parameters of a class of NLPRE with our G+D method without overparametrization, which turns out to be applicable to this example.
\endrem

\begrem
\lab{rem8}
It is clear that it is possible to define an alternative form for the $z$-dynamics \eqref{disrejz}, for instance, adding constants $a \in \rea_{>0}$ and $b \in \rea$ as
$$
\dot z(t) =-a z(t)- b Y_{\tt F}(t),
$$
and consequently redefine the matrices \eqref{disrejmat}. This modification would add some additional {\em tuning gains} to the algorithm without modifying the final result.
\endrem
%
\section{Application to CT Model Reference Adaptive Control}
\label{sec6}
%
In this section we show that, using the G+D estimator of Proposition \ref{pro2} in the classical problem of MRAC of scalar CT LTI systems it is possible to {\em remove} the standard assumption of knowledge of the high frequency gain \cite{IOASUNbook,NARANNbook,SASBODbook,TAObook}. More precisely, we consider a CT plant
\begequ
\lab{pla}
D(\bfp)y_p(t)=k_p N(\bfp)u_p(t),
\endequ
where $y_p(t) \in \rea,\;u_p(t) \in \rea$ are the plant output and input, respectively,
$$
D(\bfp)=\sum_{i=0}^{n_p}d_i \bfp^i,\quad N(\bfp)=\sum_{i=0}^{m}n_i \bfp^i,
$$
with $D(\bfp)$ and $N(\bfp)$ monic and coprime with {\em unknown} coefficients and $k_p \in \rea$ is the unknown {\em high frequency gain}.

We make the following standard assumptions regarding the plant.
\begenu
\item[{\bf A.1}] $N(\bfp)$ is a Hurwitz polynomial.
\item[{\bf A.2}] The plan order $n_p$ and relative dgree $n_p - m_p \geq 1$ are known.
\endenu

The MRAC objective is to asymptotically drive to zero the tracking error
\begequ
\lab{e}
e_T(t) = y_p(t) - y_m(t)
\endequ
where
$$
y_m(t)={ k_m \over D_m(\bfp)} r(t)
$$
with $D_m(\bfp)=\sum_{i=0}^{n_p - m_p}d^m_{i} \bfp^i$ a designer-chosen monic, Hurwitz polynomial, $k_m \in \rea$ and $r(t)$ is a bounded reference signal.

Instrumental for the proposed MRAC is the lemma below, which includes the classical direct control model reference parameterization and the {\em input error} parameterization given in \cite[Subsection 3.3.1]{SASBODbook}.

\begin{lemma}
\lab{lem1}
Consider the plant (\ref{pla}) and the tracking error (\ref{e}). There exists a vector $\theta \in \rea^{2n_p}$ such that
\begequ
\lab{traerr}
e_T(t) = { k_p \over D_m(\bfp)} [ u_p(t) - \theta^\top  \phi_{\tt PE}(t)],
\endequ
where the vector $\phi_{\tt PE}(t) \in \rea^{2n_p}$ is given by
\begin{eqnarray}
\phi_{\tt PE}(t)&=&{1 \over \lambda(\bfp)} \col \left(u_p(t),\dot u_p(t), \dots, u_p^{(n_p-2)}(t), y_p(t), \cdots \right.\nonumber \\
&&\left. \cdots, \dot y_p(t), \dots, y_p^{(n_p-2)}(t), \lambda(\bfp) y_p(t), \lambda(\bfp) r(t) \right) \nonumber \\
\lab{phipe}
\end{eqnarray}
with $\lambda(\bfp)=\Sigma_{i=0}^{n_p-1}\lambda_i \bfp^i$ a designer-chosen monic, Hurwitz polynomial. Moreover the vector $\theta$ satisfies the following input-error LRE
\begequ
\lab{inperr}
u_{\tt IE}(t)=\theta^\top \phi_{\tt IE}(t)
\endequ
where\footnote{Notice that $\phi_{N}(t)$ consists of the first $2n_p-1$ elements of  $\phi_{\tt PE}(t)$ passed through the filter ${1 \over D_m(\bfp)}$.}
\begalis{
\phi_{\tt IE}(t)&=\begmat{\phi_{N}(t)\\ {1 \over k_m}y_p(t)}\\
\phi_{N}(t) & ={1 \over D_m(\bfp)\lambda(\bfp)} \col \left(u_p(t),\dot u_p(t), \dots, u_p^{(n_p-2)}(t),  \cdots \right. \\
&\left. \cdots, y_p(t),\dot y_p(t), \dots, y_p^{(n_p-2)}(t), \lambda(\bfp) y_p(t) \right)\\
u_{\tt IE}(t)&:={1 \over D_m(\bfp)}u_p(t).
}
\qed
\end{lemma}

Motivated by \eqref{traerr}, MRAC designs are completed proposing a controller of the form
\begequ
\lab{con}
u_p(t) = \hat \theta^\top(t)  \phi_{\tt PE}(t),
\endequ
where $\hat \theta(t) \in \rea^{2n_p}$ are the estimates of the parameters $\theta$, which are generated via a parameter adaptation algorithm. As an immediate corollary of Proposition \ref{pro2} we have that using the LRE \eqref{inperr} to generate these estimates with the G+D algorithm  solves the MRAC problem requiring only the classical Assumptions {\bf A.1} and {\bf A.2} with no additional assumption on $k_p$---instead, we require identifiability of the LRE \eqref{inperr}, which is a {\em necessary} assumption for reconstruction of $\theta$.

For the sake of completeness we summarize this result in the following.

\begin{proposition}
\lab{pro6}
Consider the plant \eqref{pla} satisfying Assumptions {\bf A.1} and {\bf A.2} and LRE \eqref{inperr} with the {\em G+D interlaced} estimator
\begalis{
\dot {\hat \theta}_g(t) & = - \gamma_g(t) \phi_{\tt IE}(t)[ \phi_{\tt IE}^\top (t)\hat\theta_g(t) -   u_{\tt IE}(t)],\; \\
\dot \Phi(t)& = \gamma_g(t) \phi_{\tt IE}(t) \phi_{\tt IE}^\top(t)\Phi(t),\;\Phi(0)=I_{2n_p}\\
\dot{\hat \theta}(t) & =\gamma \Delta(t)[Y(t)-\Delta(t) \hat\theta(t)],\; \hat\theta(0)=\theta_0 \in \rea^{2n_p},
}
where $\hat\theta_g(0)=\theta_{g0} \in \rea^{2n_p}$ and  $ \gamma_g(t) \in \rea_{>0}$ and we defined the functions
\begalis{
\lab{cald}
   \cald(t) &:=I_{2n_p}-\Phi(t) \\
\Delta(t) & :=\det\{\cald(t)\}\\
Y(t) & := \adj\{\cald(t)\} (\hat\theta_g(t)-\Phi(t)\theta_{g0}).
}
If the LRE \eqref{inperr} is {\em identifiable} then
$$
\lim_{t \to \infty}\tilthe(t)=0,\;(exp).
$$
Consequently, applying the control \eqref{con} to the plant \eqref{pla} the tracking error \eqref{traerr} verifies
$$
\liminf e_T(t)=0,\;(exp).
$$
\end{proposition}

\begrem
\lab{rem9}
In \cite{SASBODbook} an estimator that uses the input-error parameterization  \eqref{inperr} and ensures global tracking is proposed. Unfortunately, this algorithm includes a parameter projection that requires, besides the knowledge of $\sign(k_p)$, and upper bound on $|k_p|$, which is essential for the proof. Moreover, it has recently been shown in \cite{BARORTacsp18} that, in the absence of the projection, input-error MRAC suffers from an instability mechanism that may give rise to unbounded trajectories---even in the simplest case of first-order plants with $r(t)=0$.
\endrem

\begrem
\lab{rem10}
As shown in \cite{BARORTacsp18} the instability mechanism of input-error MRAC is related with the lack of excitation in the regressor. This difficulty is avoided with the G+D estimator that ensures the regressor---in this case $\Delta(t)$---is PE.
\endrem

\begrem
\lab{rem11}
Input error MRAC has attracted much less attention in the adaptive control community than prediction-error MRAC. This, in spite of the fact that the former has the following significant advantages---already stressed in \cite{SASBODbook}: (i) As shown in \eqref{traerr}, in contrast to input error MRAC, the regression model used in prediction error MRAC is {\em bilinear}. To overcome this difficulty it is necessary to overparameterize the identifier excluding the possibility of parameter convergence even when PE conditions are satisfied. (ii)  The derivation of the error equation in prediction error MRAC fails if the input saturates, yielding erroneous updates in the identifier, see \cite{GOOMAY} for further discussion. These problems are conspicuous by their absence in input error MRAC.
\endrem
%
\section{Nonlinearly Parameterized Regression Equations}
\lab{sec7}
%
In this section we provide an extension of the result of Proposition \ref{pro2} to a class of NLPRE. We consider the case of CT {\em separable} NLPRE.\footnote{To avoid cluttering the notation, we restrict our presentation to the CT case, since as shown in \cite{ORTetalaut21} the extension to DT follows {\em verbatim.}} That is, NLPRE of the form
\begequ
\lab{nlplre}
  y(t)=   \phi^\top (t) \cals(\theta),
\endequ
where $\cals: \rea^q \to  \rea^p$, with $p \geq q$ is a known mapping of the vector of unknown parameters $\theta$. Similarly to \cite{BOFSLO,ORTetalaut21,TYUetal} the property that we exploit to achieve the estimation objective is {\em monotonicity}, which is defined with the following non-standard assumption.

\begenu
\item[{\bf A.3}] The mapping $\cals(\theta)$ is {\em strongly $P$-monotone} in the sense that there exists a matrix $P\in\mathbb{R}^{q\times p}$ such that
\begali{
\lab{monpro-S}
(a-b)^\top P &\left[\cals(a) - \cals(b)\right] \geq \rho|a-b|^2 >  0,\;\forall \; a,b \in \rea^q,
}
with $a \neq b$ and for some $\rho \in \rea_{>0}$.
\endenu

Notice that if $p=q$ and $P$ is positive definite Assumption {\bf A.3} reduces to the standard strong $P$-monotonicity of  $\cals(\theta)$ \cite{DEM,PAVetal}.

\begin{proposition}
\lab{pro7}
Consider the CT NLPRE (\ref{nlplre}) with  $\cals: \rea^q \to  \rea^p$ satisfying Assumption {\bf A.3}, that is \eqref{monpro-S} holds. Define the G+D interlaced estimator
\begsubequ
\lab{intestt}
\begali{
\lab{thegt}
\dot{\hat \theta}_g(t) & =\gamma_g(t) \phi(t) [y(t)-\phi^\top (t) {\hat\theta_g(t)} ],\; \hat\theta_g(0)=\theta_{g0} \in \rea^p\\
\lab{phit}
\dot { \Phi}(t)& =    \cala(t)    \Phi(t),\; \Phi(0)=I_p\\
\lab{thet}
\dot{\hat \theta}(t) & =\gamma P \Delta(t) [Y(t)-\Delta(t) \cals(\hat\theta(t)) ],\; \hat\theta(0)=\theta_0 \in \rea^q,
}
\endsubequ
where $\gamma_g(t) \in \rea_{> 0}$, $\gamma \in \rea_{> 0}$,  and we defined
\begsubequ
\lab{aydelt}
\begali{
\lab{at}
   \cala(t) &:=-\gamma_g (t) \phi(t) \phi^\top (t)\\
\lab{dt}
   \cald(t) &:=I_p-\Phi(t) \\
\lab{delt}
\Delta(t) & :=\det\{\cald(t)\}\\
\lab{yt}
Y(t) & := \adj\{\cald(t)\} [\hat\theta_g(t) -   \Phi(t)\theta_{g0}].
}
\endsubequ
If $\phi(t)$ is IE then $\lim_{t \to \infty}\tilthe(t)=0,\;(exp)$.
\end{proposition}

\begin{proof}
With some abuse of notation, define the signal
$$
\tilde \cals(t):=  \hat\theta_g(t) - \cals(\theta),
$$
whose derivative is given by
\begalis{
\dot{\tilde \cals}(t)&={\gamma_g(t) \phi(t) [y(t)-\phi^\top (t)\hat\theta_g(t)]}\\
&=-\gamma_g(t) \phi(t) \phi^\top (t) {\tilde \cals}(t)\\
&= \cala(t) {\tilde \cals}(t),
}
where we used  {\eqref{thegt}, \eqref{nlplre} and \eqref{at} to get the first, second and third identities, respectively.}   Consequently,  we get
$$
	 \tilde \cals(t)  =\Phi(t) \tilde\cals(0),
$$
which may be rewritten as the NLPRE
\begali{
\lab{keyidet}
	  \cald(t) \cals(\theta) &=  \hat\theta_g(t) -   \Phi(t)\theta_{g0},
}
where we used \eqref{dt}. Multiplying \eqref{keyidet} by   $\adj\{\cald(t)\}$ we get the following  NLPRE
\begequ
\lab{ykdelt}
Y(t) = \Delta(t) \cals(\theta),
\endequ
where we used \eqref{delt} and \eqref{yt}.
Replacing \eqref{ykdelt} in \eqref{thet} we get
\begalis{
\dot{\hat \theta}(t) & =-{ \gamma \Delta^2(t)}P[ \cals(\hat\theta(t)) - \cals(\theta)] .
}
To analyse its stability define the Lyapunov function candidate $U(\tilde \theta) := \frac{1}{2} |\tilde \theta|^2$, whose time derivative yields
\begalis{
\dot U(t)  & =  - \gamma\Delta^2(t) [ \hat \theta(t) - \theta]^\top P {[ \cals(\hat \theta(t)) - \cals(\theta)]} \\
& \leq  - \gamma \Delta^2(t) \rho | \tilde \theta(t)|^2  \\
& =  - 2\rho \gamma{\Delta}^2(t) U(t),
}
where we invoked the Assumption {\bf A.3} of strong $P$-monotonicity of $\cals(\theta)$ to get the first bound. To complete the proof, we invoke the Comparison Lemma \cite[Lemma 3.4]{KHAbook} that yields the bound
$$
U(t+t_c) \leq e^{-2 \rho\gamma \int_t^{t+t_c} \Delta^2(s)ds}U(t),
$$
which ensures $\lim_{t \to \infty}\tilthe(t)=0\;(exp)$ if  $\Delta(t)$ is PE. The latter condition follows from the assumption that $\phi(t)$ is IE and Lemma \ref{lem4}.
\end{proof}

\begrem
\lab{rem12}
The estimation procedure of Proposition \ref{pro7} differs from the one given in \cite[Proposition 2]{ORTetalaut21} in several respects. First, while the latter uses the now classical DREM estimator procedure, in the former we propose to use the new G+D estimator of Proposition \ref{pro2}. The main advantage of this modification is that we replace the assumption of $\Delta(t) \notin \call_2$, imposed in \cite{ORTetalaut21}, by the {\em strictly weaker} IE assumption of $\phi(t)$. A second fundamental difference is that the monotonicity Assumption {\bf A.3} is imposed in the present paper to the original mapping $\cals(\theta)$, this differs with \cite[Proposition 1]{ORTetalaut21} in two respects, first, a more general procedure to generate a new monotonic mapping, which involves a change of coordinates and a nonlinearity selection stage is propose in the latter. Second, the standard strict monotonicity assumption, instead of \eqref{monpro-S}, is imposed to the new mapping.
\endrem

\section{Simulations}
\label{sec8}
In this section we illustrate, via simulations, the main contributions of the paper.
\subsection{{Comparison of G+D and D+G estimators}}
\lab{subsec81}
%
To illustrate the result of Proposition \ref{pro2}, we consider in this subsection the problem of parameter estimation of a CT LTI system and choose, as an example, the system:
$$
 y_p(t) = \frac{B(\bfp)}{A(\bfp)}u_p(t) = \frac{b_1 \bfp+ b_0}{\bfp^2 + a_1 \bfp + a_0}u_p(t),
$$
where $u_p(t)\in \rea$ and $y_p(t) \in \rea$ are the input and output signals, respectively. Following the standard identification procedure \cite[Subsection 2.2]{SASBODbook} we derive the LRE \eqref{lre}  as follows
\begin{eqnarray}
    y(t)&:=&y_p(t),\;{\phi}(t) :=\begin{bmatrix}
    \frac{F(\bfp) B(\bfp)}{A(\bfp)} \\
    F(\bfp)
    \end{bmatrix} u_p(t), \nonumber \\
      F(\bfp)&:=&\frac{1}{R(\bfp)}\begin{bmatrix}
    1 \\
    \bfp \\
    \vdots \\
    \bfp^{n-1}
    \end{bmatrix}, \ \theta :=\begin{bmatrix}
    r_{0}-a_{0} \\
    \vdots \\
    r_{n-1}-a_{n-1} \\
    b_{0} \\
    \vdots \\
    b_{n-1}
    \end{bmatrix},
    \nonumber
\end{eqnarray}
with $R(\bfp)=\sum_{i=0}^{n} r_{i} \bfp^{i}, r_{n}=1$, an arbitrary Hurwitz polynomial.\\

We compare the estimation of the parameters $\theta_i$ using the G+D interlaced estimator of Proposition \ref{pro2} and the  D+G scheme based on the generation of  new LRE  presented in \cite{KORetal}, which is summarized below.
\\
Consider the scalar LRE \eqref{lre}.  Fix the constants $\lambda >0$, $g  >0$, and define the signals
\begin{eqnarray*}
\nonumber
Z(t) &=\calh[ \phi  y](t), \quad
{ \Psi}(t) =\calh[  \phi \phi^\top](t) \nonumber \\
 \caly(t) &= \adj\{ \Psi(t)\} Z(t), \quad  \bar \Delta(t) =\det\{ \Psi(t)\},
\end{eqnarray*}
where $\calh[u]= {g \over \bfp + \lambda}[u](t)$ is a linear filter. In  \cite{KORetal} it is shown that if $\phi $ is IE then, $\bar \Delta$ is IE and the $q$ scalar LREs
$$
 \caly_i(t)=  \bar \Delta(t)    \theta_i,\;i \in \bar q,
$$
hold. Now,  define the dynamic extension.\footnote{To simplify the notation we omit the subindex $i$.}
\begalis{
\dot z(t) & =   -k z(t) + k {\bf \bar \Phi}_1(t) \caly(t) ,\; z(0)=0 \\	
	 \dot \zeta(t) & =  \bar A(t)  \zeta(t) +  \bar b(t),\;  \zeta(0)=\col(0,0) 	 \\
\dot  {\bar {\bf \Phi}}(t) & =  \bar A(t) {\bf \bar \Phi(t)},\;  {\bf \bar \Phi}(0)=\col(1,0),
}
and
\begin{eqnarray}
\bar A(t) &:=&\begmat{0 & -k \bar \Delta(t) {\bf \bar \Phi}_1(t) \\  k\bar \Delta(t) {\bf \bar \Phi}_1(t)  &  -\tilde V(t)},\;  \nonumber \\
 \bar b(t)&:=&\begmat{- k \bar \Delta(t) {\bf \bar \Phi}_1(t) z(t)\\   [\tilde V(t)-k]  z(t)}, \nonumber
\end{eqnarray}
where
$$
   \tilde V(\bf \bar \Phi) := \frac{1}{2}\left({\bf \bar \Phi}_1^2 + {\bf \bar \Phi}_2^2\right)-\beta,
$$
with $k >0$ and $\beta>\frac{1}{2}$. Then, the new LRE
$$
{\bar {\bf Y}}(t)= {\bf \bar \Phi}_{2}(t) \theta,
$$
holds with
$$
{\bar {\bf Y}}(t):= z(t)-  \zeta_{2}(t).
$$
Moreover, ${\bf \bar \Phi}_{2}(t)$ is PE and $z(t)$, $\zeta(t)$, ${\bf \bar \Phi}(t)$ are bounded. Hence, using the standard gradient descent adaptation
$$
	\dot{\hat{\theta}}(t)= \kappa {\bf { \bar \Phi}} (t)\left({\bf \bar Y} (t)-{\bf \bar \Phi} (t)\hat{\theta}(t)\right),\;\kappa > 0,
$$
we get {\em exponential parameter convergence}. For further details of the D+G scheme see   \cite[Propositions 1 and 2]{KORetal}.

To carry out the simulations, we use the system studied in  \cite[Section 5]{aranovskiy2019parameter}, that is $(a_0,a_1,b_0,b_1)=(2,1,1,2)$ and choosing  $r_1 = 20$ and $t_0 = 100$. This yields $ \theta = \col(98,19,1,2)$ and for both estimators we propose an input signal that is not sufficiently rich, but generate a regressor $\phi(t)$ which is IE, namely
$$
    u_{p}(t) = e^{-2t}+e^{-1.5t}.
$$

From Proposition \ref{pro2} it is clear that the G+D estimator has only two tuning gains $\gamma$ and $\gamma_g$ that, as discussed in Remark \ref{rem2}, have a clear role in the transient behavior. From the material above we see that the D+G scheme has  $ \lambda, g, k, \beta$ and $\kappa$,  whose impact on the transitory is rather obscure.  Simulation experience has shown that tuning the gains of the D+G estimator is a hard task and  a bad selection can have profoundly adverse effects on the behavior of the estimation. On the other hand,  the tuning of the G+D estimator is relatively straightforward. To illustrate these facts,  we present below some comparative simulations of both schemes. We focus on two important parameters of both schemes, which are,  $\gamma_g$ and $\beta$, where both play a central role in the generation of the ``exciting signals'' in the {\it regressors} of the new LREs. Hence, we   fixed the gains of the D+G scheme to  $\lambda =g=1,k=0.4$ and $\kappa=10$, while for the G+D estimator we use $\gamma=200$  and  propose the following different values for $\beta$  and $\gamma_g$
\begin{eqnarray*}
\beta&=&\{0.65, \; 0.8, 1.1, 1.5 \} \nonumber \\
\gamma_g&=&\{2500, \; 2900, \; 3300, \;3800\}.
\end{eqnarray*}
Using the same set of gains for all estimated parameters, all  initial conditions $\hat{\theta}_i(0)=0$ in both estimators  and $\theta_{g_0}=\col(0.4,0.2,0,0.5)$ for the G+D one.

\begin{figure}[http]
\centering
\includegraphics[width=0.52\textwidth]{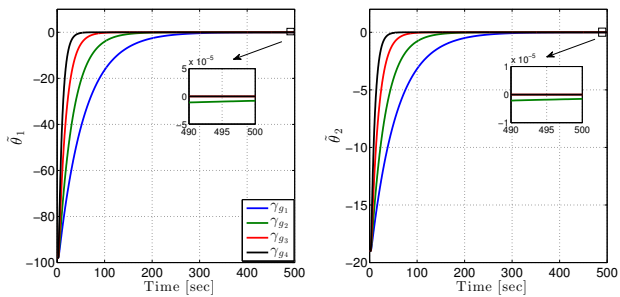}
\caption{Transient behavior of $ \tilde \theta_1(t)$ and $\tilde \theta_2(t)$ using the G+D scheme }
\label{LTI1}
\end{figure}

\begin{figure}[http]
\centering
\includegraphics[width=0.52\textwidth]{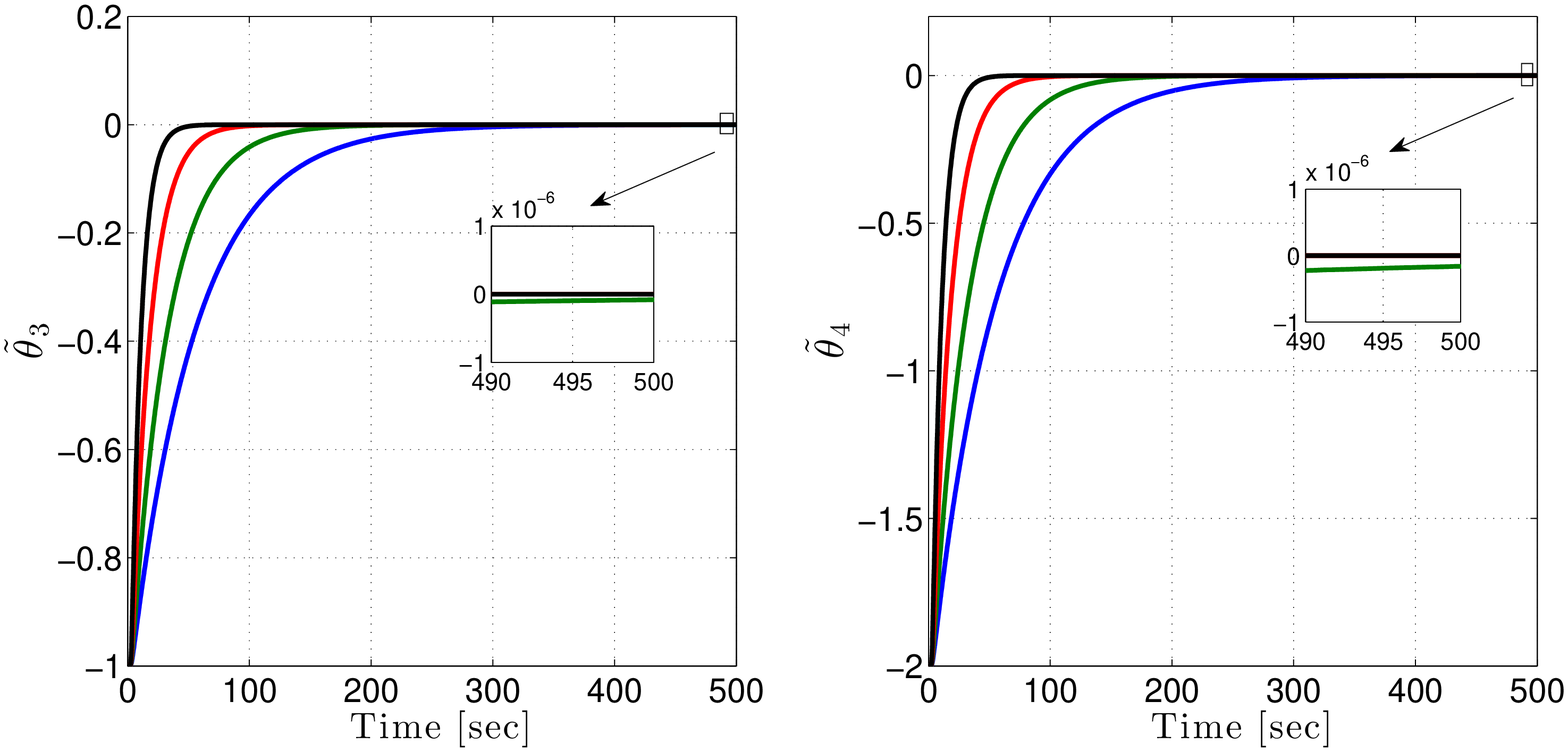}
\caption{Transient behavior of $ \tilde \theta_3(t)$ and $\tilde \theta_4(t)$ using the G+D scheme }
\label{LTI2}
\end{figure}

\begin{figure}[http]
\centering
\includegraphics[width=0.52\textwidth]{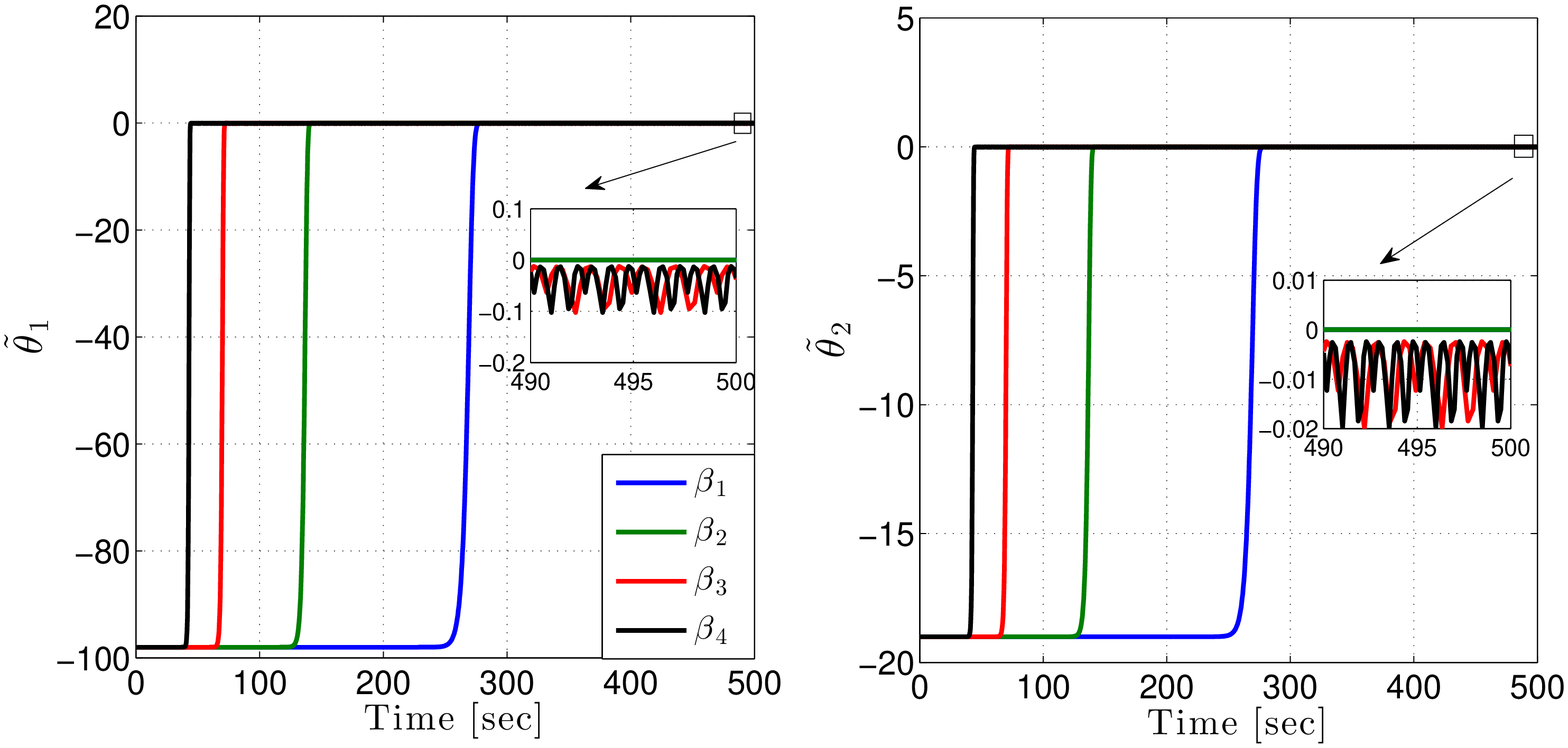}
\caption{Transien behavior of $ \tilde \theta_1(t)$ and $\tilde \theta_2(t)$ using the D+G scheme}
\label{LTI3}
\end{figure}

\begin{figure}[http]
\centering
\includegraphics[width=0.52\textwidth]{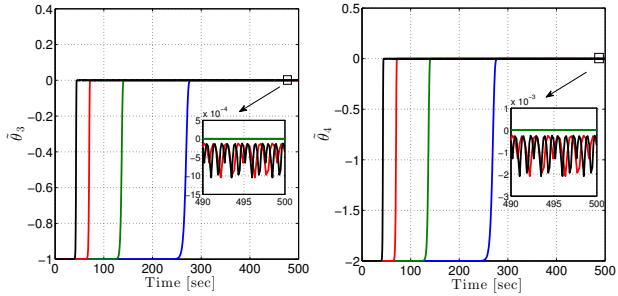}
\caption{Transient behavior of  $ \tilde \theta_3(t)$ and $\tilde \theta_4(t)$ using the D+G scheme }
\label{LTI4}
\end{figure}

The simulation results, which corroborate the claims above, are shown in Figs. \ref{LTI1}--\ref{LTI4}, where we depict the behavior of each $\tilde \theta_i(t)$ for each value of $\beta$ and $\gamma_{g}$, distinguishing them by the color in the Figures. It is appreciated that the convergence of the G+D has a clear monotonic behavior with respect to  $\gamma_g$. The convergence of the D+G one is also faster as $\beta$ increases but this introduces a  ``dead-time'' in the response. We also see  from Figs. \ref{LTI3} and \ref{LTI4}  that if $\beta>1$,  the D+G scheme generates an unusual oscillatory behavior around zero of non-negligible amplitude---see the difference in the scales of the boxed regions. The reason for the appearance of both undesirable effects is  not clear and does not follow from the theoretical analysis in \cite{KORetal}.

\subsection{{Application of G+D MRAC to Rohrs' Examples}}
\lab{subsec82}
%
In this subsection, we evaluate the performance of the G+D MRAC of Proposition \ref{pro6}. In particular, we consider the scenarios of \cite{ROHetal}, which are widely used as benchmarks to study the robustness of adaptive controllers {\em vis-\`a-vis} unmodeled dynamics and noise.

We consider a first-order plant
$$
y_p(t) = {2\over \bfp+1}[u_p](t),
$$
and a reference model
$$
y_m(t) = {3\over \bfp + 3}[r](t).
$$
In this case the ideal controller gains are $\theta=\col(1.5,-1)$.  We adopt the reference signals proposed in \cite{ROHetal}, that is $r_1(t) = 0.3 + 18.5 \sin (16.1t)$ and $r_2(t) = 2.$  According to \cite{SASBODbook}, the reference $r_1(t)$ is ``sufficiently rich'' for a plant with two unknown parameters, thus the associated regressor of prediction error MRAC \eqref{phipe} satisfies the PE condition. On the other hand, for the reference $r_2(t)$, the PE condition is not satisfied and parameter convergence cannot be guaranteed.

\begin{figure}
 \centering
  {
    \includegraphics[width=0.23\textwidth]{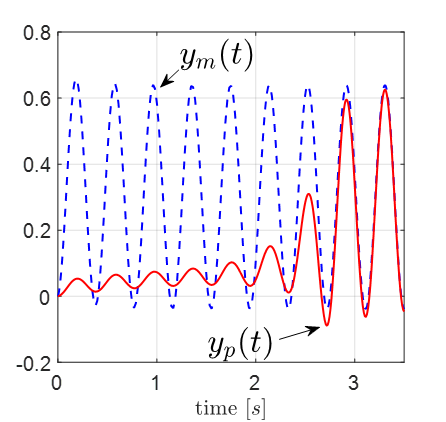}}
 {
   \label{f:tigre}
    \includegraphics[width=0.23\textwidth]{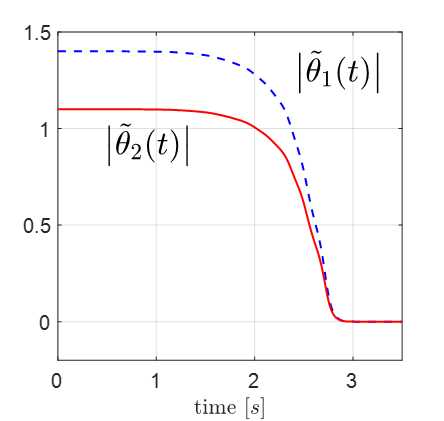}}
       \caption{Simulation results of the G+D  MRAC  with the reference $r_1(t)$ in the ideal case }
    \label{fig:mrac1}
\end{figure}

\begin{figure}
 \centering
  {
    \includegraphics[width=0.23\textwidth]{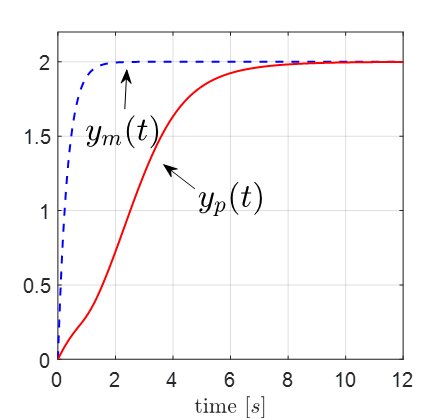}}
 {
   \label{f:tigre}
    \includegraphics[width=0.23\textwidth]{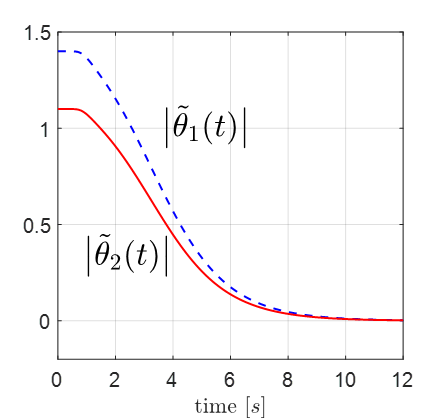}}
 \caption{Simulation results of the G+D  MRAC with the reference $r_2(t)$ in the ideal case}
    \label{fig:mrac2}
\end{figure}

First, we simulated the G+D MRAC for the ideal case in the absence of unmodeled dynamics. The initial conditions are set as $\hat \theta(0) = [0.1, 0.1]^\top$ and all the others are selected as zero. The gains are chosen as $\gamma_g =200 $ and $\gamma =100$, with the simulation results shown in Figs. \ref{fig:mrac1} and \ref{fig:mrac2}. In both cases, we get satisfactory tracking performance, and the parameter estimation errors exponentially converge to zero even for the non sufficiently rich reference $r_2(t)$.

\begin{figure}
 \centering
  {
    \includegraphics[width=0.23\textwidth]{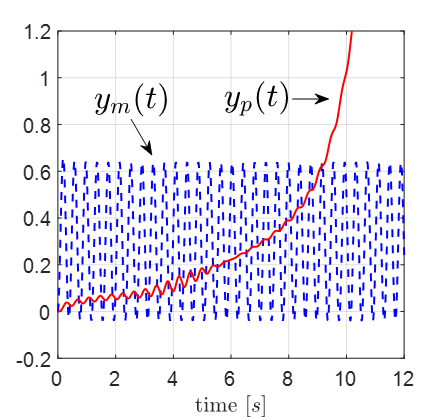}}
 {
   \label{f:tigre}
    \includegraphics[width=0.23\textwidth]{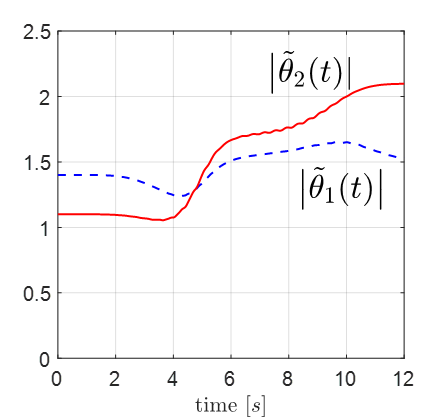}}
 \caption{Simulation results of the G+D  MRAC  with unmodelled dynamics, the reference $r_1(t)$ and constant gain $\gamma_g$}
    \label{fig:mrac3}
\end{figure}

\begin{figure}
 \centering
  {
    \includegraphics[width=0.23\textwidth]{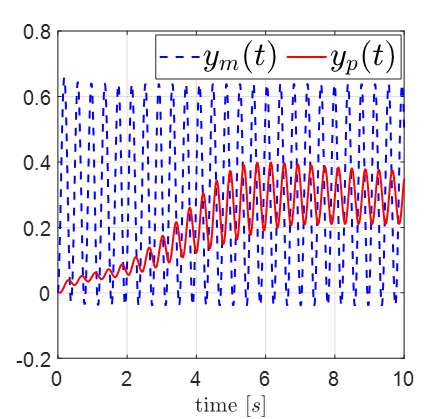}}
 {
   \label{f:tigre}
    \includegraphics[width=0.23\textwidth]{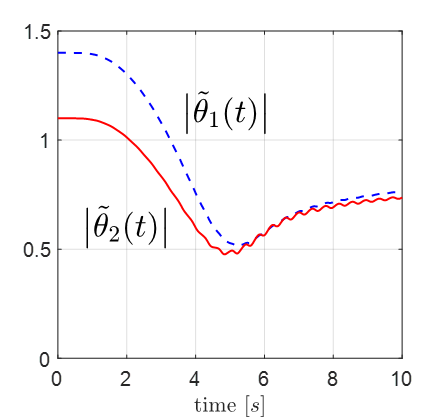}}
\caption{Simulation results of the G+D  MRAC  with unmodelled dynamics, the reference $r_1(t)$ and time-varying gain $\gamma_g(t)$}
    \label{fig:mrac30}
\end{figure}

As shown in  \cite{ROHetal}---see also \cite[Subsection 5.2]{SASBODbook}---in the presence of unmodeled dynamics of the form
$$
y_p(t) = {2\over \bfp + 1 } \cdot {229 \over \bfp^2 + 30\bfp + 229}[u_p](t),
$$
 prediction error MRAC will diverge for both reference signals. To assess the robust performance of G+D  MRAC and verify the robustification claims of Proposition \ref{pro4} we simulated the estimator, for the PE reference $r_1(t)$,  with a constant gain $\gamma_g = 100$, and a time-varying gain $\gamma_g(t) = {100\over 0.1 + t^2}$, both with $\gamma =1000$. From Fig. \ref{fig:mrac3}, we see that  the constant gain adaptive controller is unstable.  On the other hand, using a time-varying gain $\gamma_g(t)$ guarantees signal boundedness of the closed-loop with very good parameter estimation as shown in  Fig. \ref{fig:mrac30}. It should be underscored that, due to the presence of the unmodeled dynamics, parameter convergence to the ideal plant values is no guarantee for stability.

\begin{figure}
 \centering
  {
    \includegraphics[width=0.23\textwidth]{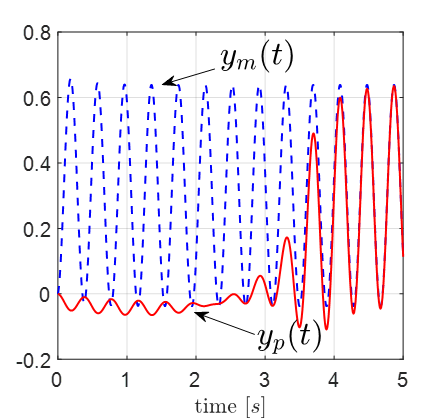}}
 {
   \label{f:tigre}
    \includegraphics[width=0.23\textwidth]{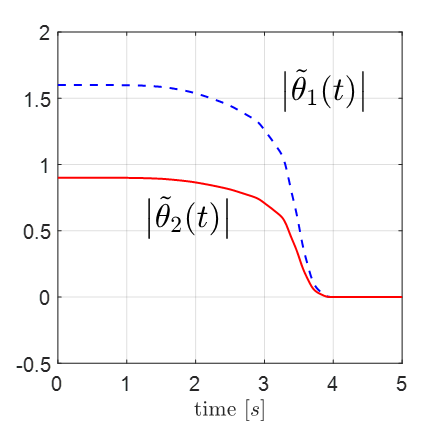}}
    \caption{Simulation results for MRAC with the reference $r_1$ and unknown sign of the high-frequency gain}
    \label{fig:mrac4}
\end{figure}

Finally, we considered the case of unknown sign of the high-frequency gain $k_p$. To this end, we test the same controller with $\gamma_g =200$ and $\gamma=100$, but applying to the plant
$$
y_p(t) = {-2\over \bfp +1 }[u_p](t).
$$
The simulation results, given in Fig. \ref{fig:mrac4}, show that the proposed method still guarantees exponential convergence of both the state and estimation errors.

\subsection{{Disturbance rejection with a NLPRE}}
\lab{subsec84}
%
In this section, simulations of the parameter estimation  for the rejection of sinuoidal disturbances problem formulated in Section \ref{subsec52} are presented. First, we recall that using Proposition \ref{pro5} we can generate from the {\it scalar} perturbed LREs  \eqref{disLRE} {\em unperturbed} LREs of the form
\begin{equation}
\label{LREaux}
{\mathcal Y}_2(t)= {\tilde \Omega}^\top (t)  \begmat{\theta \\ \frac{\theta}{\omega^2} \\ \frac{1}{\omega^2}}
\end{equation}
with
\begalis{
{\mathcal Y}_2(t) &= z(t)-r_2(t)\\
\tilde \Omega(t)&=\begmat{(\Phi_\xi)_{2,1}(t) & \Omega_{2,1}(t)  & \Omega_{2,2}(t)}
}
of Proposition \ref{pro5}. This LRE may be seen as  a NLPRE of the form \eqref{nlplre} with $y(t)= {\mathcal Y}_2(t)$, $\phi(t)= \tilde \Omega(t)$ and
\begequ
\lab{calsthe}
\mathcal{S}\Big(\theta,\frac{1}{\omega^2}\Big) = \begmat{\theta \\ \frac{\theta}{\omega^2} \\ \frac{1}{\omega^2}},
\endequ
which is  nonlinearly parameterized with respect to the physical parameters $(\theta,\omega)$. We make now the important observation that choosing
\[
P=\begmat{1 & 0& 0 \\ 0& 0& 1}
\]
we verify {\bf A.3}, that is, $P\nabla\cals + (P\nabla\cals)^\top = 2 I_2$, so that, $\mathcal{S}\Big(\theta,\frac{1}{\omega^2}\Big)$ is strongly $P$-monotone. Hence, it is possible to apply the estimation algorithm for NLPRE of Proposition \ref{pro7}.

Towards this end, we notice that this choice of $P$ allows us to consider as unknowns only the first and third elements of the mapping $\mathcal{S}\Big(\theta,\frac{1}{\omega^2}\Big)$. For future refence, we rename them
$$
\theta_1:=\theta,\;\theta_2:=\frac{1}{\omega^2}.
$$
We present now simulations of the estimator im Proposition \ref{pro7} to estimate this unknown parameters.

\begin{figure}
\centering
\includegraphics[width=0.53\textwidth]{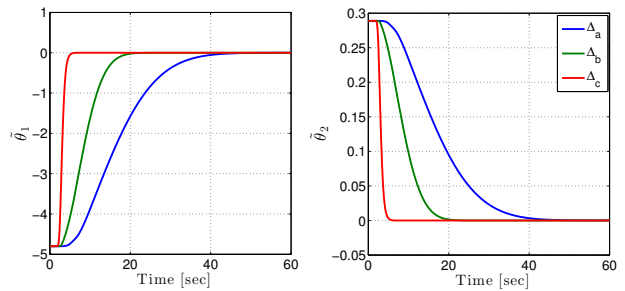}
\caption{Transient behavior of  $\tilde \theta_1(t)$ and $\tilde \theta_2(t)$}
\label{sim_nplredis}
\end{figure}

\begin{figure}
\centering
\includegraphics[width=0.53\textwidth]{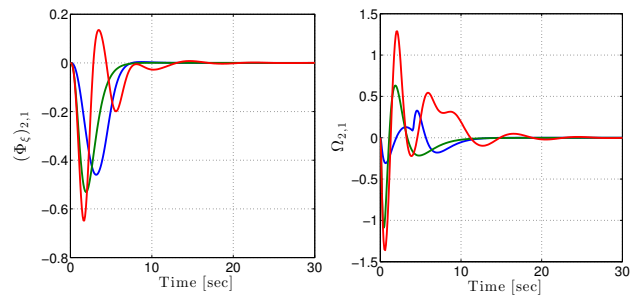}
\caption{Transient behavior of  $(\Phi_{\xi})_{2,1}(t)$ and $\Omega_{2,1}(t)$}
\label{sim_signewLRE1}
\end{figure}

\begin{figure}
\centering
\includegraphics[width=0.53\textwidth]{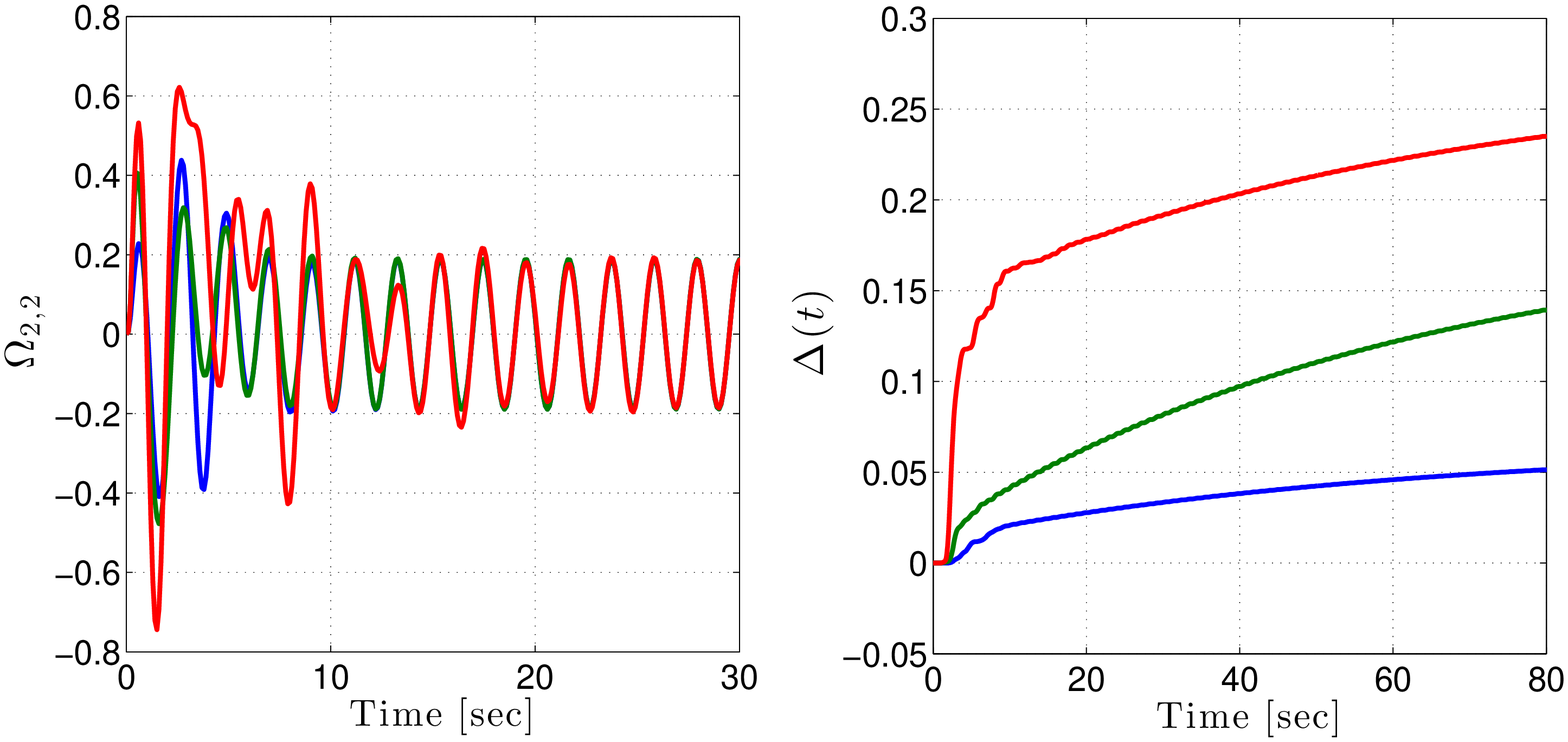}
\caption{Transient behavior of   $\Omega_{2,2}(t)$ and $\Delta(t)$ of Proposition \ref{pro7}}
\label{sim_signewLRE2}
\end{figure}

We fix the disturbance $\xi=0.5 \sin(\omega t + 0.4)$ with {\it unknown} frequency $\omega=5$ and $\theta=5$. The initial conditions that were used in all simulations are $\theta_{g_0}= \col(0.4,\;0.2,\;0.5)$ and $\theta_0=\col(0.2, \;0.4)$. Besides, we set the tuning gains to $\lambda=1$, $\gamma_g= 0.9$ and $\gamma=150$. To evaluate the effect of the richness content of the input signal on the  performance of the estimator of Proposition \ref{pro7} we consider the following  three different signals $\Delta(t)$, namely:
  \begin{eqnarray*}
\begin{aligned}
    \Delta_a(t)& =
    \left\{
    \begin{aligned}
         1 & &t \in[0,4]
         \\
         0 & & t > \mbox{4}
    \end{aligned}
    \right. \\
    \Delta_b(t) & = {1\over 0.2 + t^2}. \\
     \Delta_c(t) & = 5 \exp^{-0.2 t} \cos\left(\frac{\pi}{4} t \right).
\end{aligned}
\end{eqnarray*}
Even though the three signals are not  PE and belong to ${\mathcal L}_2$,  they are IE, thus the estimation (without overparametrization) of $\theta_1$ and $\theta_2$ is  guaranteed using the estimator \eqref{intestt}. Also, it is clear that the richness content of the signal above increases from the first to the last one. Fig. \ref{sim_nplredis} corroborates this fact, where the line color distinguishes the signal $\Delta_{(\cdot)}(t)$ that is used. On the other hand, Figs. \ref{sim_signewLRE1} and  \ref{sim_signewLRE2}  show the transient behavior of the signals $(\Phi_{\xi})_{2,1}(t)$, $\Omega_{2,1}(t)$ and $\Omega_{2,2}(t)$  of the new LRE  \eqref{disrejlre} and the signal $\Delta(t)$ of the interlaced estimator defined in \eqref{delt} in Proposition \ref{pro7}. The latter plot clearly confirms our claim regarding the richness of the signals $\Delta_{(\cdot)}(t)$.

%
%
\section{Conclusions and Future Research}
\label{sec9}
%
In this paper we have provided a solution to the problem of designing an on-line,  estimator that ensures GES of the PEE under the weakest assumption that the LRE is {identifiable}. Moreover, we have shown that, imposing a constraint on the adaptation gain of the first estimator, we prove that the scheme is robust to external disturbances and (not necessarily slow) parameter variations. We also proposed a variation of this estimator that rejects sinusoidal disturbances with unknown internal model and shown that the procedure is applicable to a well-defined class of NLPRE. Finally, we showed that, applying the proposed estimator in the MRAC problem, ensures relaxes the assumption of known sign of the high frequency gain.

Our current research efforts are oriented in the following directions.
\begite
\item Application of the G+D estimator to the problem of {\em state observation} of state-affine systems as done in \cite{ORTetalaut}, from which is clear that the  conditions for convergence will be relaxed and the robustness properties improved. In particular, we are interested in the case of time-varying systems with unknown parameters as done in \cite{BOBetalijc21}.
\item Extend the material of Section \ref{sec6} to address other issues arising in standard MRAC. For instance robustness to unmodeled dynamics and the required prior knowledge for the multivariable case \cite{GERetalsysid,TAO}. {The simulation results of Section \ref{sec8}, that reveal some robustness of the new scheme with respect to the classical counterexamples of \cite{ROHetal}, being quite encouraging. However, a deeper understanding of the instability mechanisms, partially revealed in \cite{BARORTacsp18}, is required.}
\item Extend the disturbance rejection result of Proposition \ref{pro5} to the case of multiple frequencies. We have available a solution for two frequencies but the generalization to more frequencies is still to be worked out.
\item Proceed with the comparative study of the proposed G+D estimator and the D+G one proposed in \cite{KORetal}. {As shown in Section \ref{sec8} the behavior of G+D is ``monotonic" with respect to the tuning gains $\gamma$ and $\gamma_g$. On the other hand, simulation evidence has shown that D+G has a more ``erratic" dependence on $\lambda,g,\alpha,\beta,k$ and $\kappa$, hence the commissioning procedure of the former is ``easier". In any case, a better understanding of their similarities/differences is needed,}
\item In \cite{YIetalcdc21} a procedure to ``mix" the estimates $\hat \theta_g(t)$ and $\hat \theta(t)$ that preserves the main properties of Proposition \ref{pro2} is proposed. The advantages of this new modification are yet to be clarified.
\endite

\begcen
{\bf  Acknowledgements}
\endcen
The authors are greatful to S. Aranovskiy for help in the proof of Lemma \ref{lem4} and the derivations of Subsection \ref{subsec52}.

\appendices

\section{Preliminary Lemmata}
%
In this appendix we prove that $\phi$ in IE implies that the signal $\Delta$ generated according to the construction of Proposition \ref{pro2} is PE. The CT and DT cases are given in Lemmas \ref{lem3} and \ref{lem5}, respectively.

Instrumental for the establishment of the CT claim is the following result from  \cite[Lemma~1]{EFIFRA}.

\begin{lemma}
\label{lem2}
	Let $w(t)\in\mathbb{R}^q$ be the solution of
	\begequ
	\lab{dotz}
		\dot{w}(t) = -{\gamma_g(t)} \phi(t) \phi^\top(t)w(t), \;\quad w(0)=w_0\in\mathbb{R}^q,
	\endequ
with $\phi(t)$ being IE  {and $\gamma_g(t) \in \rea_{>0}$} being {\em continuous} and bounded. Define $W(w):= \frac{1}{2}|w|^2$. Then, it holds that
	\begequ
	\lab{ineefi}
		W(0) - W(t_c) \ge \frac{\bar \gamma_g C_c}{1+\bar \gamma_g^2 t_c^2{\phi_M^2}} W(0),
	\endequ
where ${\phi_M}:=\max_{t\in[0,t_c]}|\phi(t)|^2$, {$\bar \gamma_g:=\min_{t\in[0,t_c]}\gamma_g(t)$}, and $t_c,C_c$ are given in Definition \ref{def1}.
\end{lemma}

\begin{lemma}
\lab{lem3}[CT case]
	Let $\Phi(t)$ be the solution of \eqref{phi}, \eqref{cala} with $\phi(t)$ bounded {for $t\in[0,t_c]$} and IE. Then, there exists $\epsilon\in (0,1]$ such that
	\[
		{|\Delta(t)|} \ge \epsilon^q
	\]
	for all $t\ge t_c$. Consequently, $\Delta(t)$ is PE.
\end{lemma}
\begin{proof}
	Consider the LTV system \eqref{dotz} for an arbitrary nonzero $w_0$, whose fundamental matrix is $\Phi(t)$. Hence, $w(t) = \Phi(t)w_0$ and
	\[
		W(t) = \frac{1}{2}|w(t)|^2= \frac{1}{2} w_0^\top\Phi^\top(t) \Phi(t) w_0
	\]
	and
	\[
		W(0) - W(t_c)  = \frac{1}{2} w_0^\top\left[I_n - \Phi^\top(t_c) \Phi(t_c)\right] w_0.
	\]
	Invoking \eqref{ineefi}, we get the bound
	\[
		W(t_c)=\frac{1}{2}|\Phi(t_c)w_0|^2 \le \frac{1}{2}\left(1-\frac{\bar \gamma_g C_c}{1+\bar \gamma_g^2 t_c^2{\phi_M^2}}\right)|w_0|^2.
	\]	
Since $W(t)$ is non-increasing, it immediately follows that
\[
|\Phi(t)w_0|^2 \le \left(1-\frac{\bar \gamma_g C_c}{1+\bar \gamma_g^2 t_c^2\phi_M^2}\right)|w_0|^2\,,\quad \forall t\ge t_c\,.
\]
Further, as $w_0$ is an arbitrary nonzero vector, it follows  that  the spectral radius of $\Phi(t)$ satisfies
	\begequ
	\lab{lammaxphi}
		{\rho}\left\{\Phi(t)\right\} \le \sqrt{1-\frac{\bar \gamma_g C_c}{1+\bar \gamma_g^2 t_c^2{\phi_M^2}}} < 1,\quad \forall t\ge t_c
	\endequ
where we used the fact that  ${\phi_M}t_c \ge C_c$---which follows from Definition \ref{def1}---and thus $\frac{\bar \gamma_g C_c}{1+\bar \gamma_g^2 t_c^2{\phi_M^2}} \in (0,1)$. Consequently, each eigenvalue $\lambda_i$, $i\in\bar q$ of $I_q - \Phi(t)$ satisfies
	\[
		|\lambda_i\{I_q - \Phi(t)\}| \ge 1-\sqrt{1-\frac{\bar \gamma_g C_c}{1+\bar \gamma_g^2 t_c^2{\phi_M^2}}}=:\epsilon ,\quad \forall t\ge t_c.
	\]
 The proof is completed recalling that $\Delta(t)=\det\{I_q - \Phi(t)\}$ and the fact that a scalar function whose lower bound converges to a non zero value is PE.
\end{proof}

\medskip
Instrumental for the establishment of the DT claim is the following result.

\begin{lemma}
\label{lem4}
	Let $w(k)\in\mathbb{R}^q$ be the solution of
	\begequ
	\lab{z_k+1}
		{w}(k+1) = \left(I_q-\frac{\phi(k) \phi^\top(k)}{\gamma_g(k)+|\phi(k)|^2}\right)w(k), \quad w(0)=w_0\,,
	\endequ
with $\phi(k)$ being IE and $\gamma_g(k) \in \rea_{>0}$. Then, there exists $\alpha_0\in(0,1)$ such that
	\begequ	\label{eq:27}
		|w(k)| \le \alpha_0|w_0|\,,\; \forall k\geq k_d
	\endequ
with $k_d$ given in Definition \ref{def1}.
\end{lemma}
\begin{proof}
It is observed that for all $\bar k\geq k_d$,
\begin{align*}
&\sum_{k=0}^{\bar k}\frac{\phi(k) \phi^\top(k)}{\gamma_g(k)+|\phi(k)|^2} \geq\sum_{k=0}^{k_d}\frac{\phi(k) \phi^\top(k)}{\gamma_g(k)+|\phi(k)|^2} \geq \cdots \\
&\cdots  \sum_{k=0}^{k_d}\frac{\phi(k) \phi^\top(k)}{\gamma_g(k)+{\phi_{M_k}}} \geq \frac{C_d}{\bar \gamma_g+{\phi_{M_k}}}I_q >0\,,
\end{align*}
where the second inequality is obtained by defining {${\phi_{M_k}}:=\max_{k\in[0,k_d]}\|\phi(k)\|^2$} and the third is obtained by using the IE condition of $\phi(k)$ and defining $\bar \gamma_g :=\max_{k\in[0,k_d]}\gamma_g(k)$.
Therefore, by recalling  \cite[Proposition 3.3]{TAObook},  \eqref{eq:27} can be concluded.
\end{proof}

\begin{lemma}
\lab{lem5}[DT case]
If  $\phi(k)$ is IE, then there exists $\delta _d  \in \rea_{> 0}$ such that
$$
|\Delta(k)|\geq \delta _d\,,\quad \forall k \geq k_d.
$$
Consequently, $\Delta(k)$ is PE.
\end{lemma}
\begin{proof}
Observe that
$$
h^\top D(k) h = h^\top[I_q-\Phi(k)]h >0,\; \forall h \in \rea^q \setminus \{0\}\;\Rightarrow \;  |\Delta(k)|  \in \rea_{> 0}
$$
and the left hand-side inequality holds if
$$
  |\Phi(k)h| < |h|\,,\quad \forall h\in\mathbb{R}^q\setminus \{0\}.
$$
Whence, we will prove the lemma showing that, if $\phi(k)$ is IE, there holds
\begin{equation}\label{eq:epsilon}
  |\Phi(k)h| \leq \alpha_0|h|\,,\quad \forall k \geq k_d, h\in\mathbb{R}^q\,
\end{equation}
with $\alpha_0\in(0,1)$ given in \eqref{eq:27}. It is also noted that $\Phi(k)$ is the fundamental matrix of the system \eqref{z_k+1}, which implies $w(k)=\Phi(k)h$ with $w_0=h$. In this way, the proof reduces to show
\[
|w(k)| \leq \alpha_0|w_0| \,,\quad \forall k \geq k_d, w_0\in\mathbb{R}^q\,.
\]
which clearly is  true by Lemma \ref{lem4}.

Therefore, there holds \eqref{eq:epsilon} and thus $|\Delta(k)|\geq (1-\alpha_0)^q$ for all $k \geq k_d$, completing the proof.
\end{proof}
\newpage

\section{List of Acronyms}
%
{
\begin{table}[h]
	\centering
	\label{tab:2}
	\renewcommand\arraystretch{1.6}
	\begin{tabular}{l|r}
		\hline\hline
		BIBO & Bounded-input bounded-output \\
		CT & Continuous-time \\
		DREM  &  Dynamic regressor extension and mixing \\
		DT & Discrete-time \\
		D+G & DREM plus GPEBO \\
		GPEBO & Generalized parameter estimation based observer\\
		GES & Global exponential stability\\
		G+D & GPEBO plus DREM\\
		IE & Interval excitation \\	
		KP & Key problem\\ 	
		LRE & Linear regressor equation \\
		LTI &  Linear time-invariant  \\
		LTV &  Linear time-variant  \\
		MRAC & Model reference adaptive control\\
		NLPRE & Nonlinearly parameterized regressor equations\\
		PE &  Persistent excitation  \\
		PEE &  Parameter error equations  \\
		\hline\hline
	\end{tabular}
\end{table}
}


\begin{thebibliography}{00}

\bibitem{ARAetalpe}
S. Aranovskiy, A. Bobtsov, A. Pyrkin, R. Ortega and A. Chaillet, Flux and position observer of permanent magnet synchronous motors with relaxed persistency of excitation conditions, {\it IFAC-PapersOnLine}, vol. 48, no. 11, pp. 301-306, 2015.

\bibitem{ARAetaltac}
S. Aranovskiy, A. Bobtsov, R. Ortega and A. Pyrkin, Performance enhancement of parameter estimators via dynamic regressor extension and mixing,  \TAC, vol. 62, pp. 3546-3550, 2017. (See also {\tt arXiv:1509.02763} for an extended version.)


%
\bibitem{aranovskiy2019parameter}
S. Aranovskiy, A. Belov, R. Ortega, N. Barabanov and A. Bobtsov,  Parameter identification of linear time--invariant systems using dynamic regressor extension and mixing. {\it International Journal of Adaptive Control and Signal Processing}, vol. 33, no. 6, pp. 1016-1030, 2019.



\bibitem{BARORT}
N. Barabanov and R. Ortega, On global asymptotic stability of $\dot x = \phi(t)\phi^\top (t)x$ with $\phi(t)$ bounded and not persistently exciting, {\em Systems and Control Letters}, vol. 109, pp. 24-27, 2017.

\bibitem{BARORTacsp18}
N. Barabanov and R. Ortega, On the need of projections in input-error model reference adaptive control, {\it Int. J. on Adaptive Control and Signal Processing}, vol. 32, vo. 3, pp. 403-411, 2018.

\bibitem{BELetalsysid}
A. Belov, R. Ortega and A. Bobtsov, Guaranteed performance adaptive identification scheme of discrete-time systems using dynamic regressor extension and mixing, {\em 18th IFAC Symposium on System Identification, (SYSID 2018)}, Stockholm, Sweden, July 9-11, 2018.

\bibitem{BOBetal}
A. Bobtsov, B. Yi,  R. Ortega and A. Astolfi, Generation of new exciting regressors for consistent on-line estimation of a scalar parameter, \TAC, (to be published), 2021. ({\tt arXiv:2104.02210}.)

\bibitem{BOBetalijc21}
A. Bobtsov, R. Ortega, B. Yi and  N. Nikolayev,  Adaptive state estimation of state-affine systems with unknown time-varying parameters, \IJC,  ({\tt DOI:10.1080}, {\tt 00207179.2021.1913647, Article ID:TCON 1913647}), 2021.
%
\bibitem{BOFSLO}
N. M. Boffi and J.-J. Slotine, Higher-order algorithms and implicit regularization for nonlinearly parameterized adaptive control, {\em MIT Int. Report}, Mar. 2020. ({\tt arXiv:1912.13154v3}).


\bibitem{BRUPROKUT}
S. Brunton, J. Proctor and J, Kutz, Discovering governing equations from data by sparse identification of nonlinear dynamical systems, {\em Proceedings of the National Academy of Science}s, vol. 113, no. 15, pp. 3932-3937, 2016.

\bibitem{CHOetal}
G. Chowdhary, T. Yucelen, M. Muhlegg and E. N. Johnson, Concurrent learning adaptive control of linear systems with exponentially convergent bounds, {\em International Journal of Adaptive Control and Signal Processing}, vol. 27, no. 4, pp. 280-301, 2013.


\bibitem{DEM}
B. P. Demidovich, Dissipativity of nonlinear systems of differential equations, {\it Vestnik Moscow State University, Ser. Mat. Mekh., Part I-6}, (1961) pp. 19-27; {\it Part II-1}, (1962), pp. 3-8, (in Russian).

\bibitem{EFIFRA}
D. Efimov and A. Fradkov,  Design of impulsive adaptive observers for improvement of persistency of excitation, \emph{International Journal of Adaptive Control and Signal Processing}, vol. 29, no. 66, pp. 765-782, 2015.

\bibitem{EGA}
B. Egardt, {\em Stability of Adaptive Controllers}, New York: Springer-Verlag, 1979.


\bibitem{GERetalsysid}
D. Gerasimov, R. Ortega and V. Nikiforov, Adaptive control of multivariable systems with reduced knowledge of high frequency gain: Application of dynamic regressor extension and mixing estimators, {\em 18th IFAC Symposium on System Identification, (SYSID 2018)}, Stockholm, Sweden, July 9-11, 2018.

\bibitem{GOOSINbook}
G. Goodwin and K. Sin, {\em Adaptive Filtering Prediction and Control}, Prentice-Hall, 1984.

\bibitem{GOOMAY}
G. Goodwin and D. Mayne, A parameter estimator perspective of continuous-time model reference adaptive control, \AUT, vol. 23, no. 1, pp. 57-70, 1987.

\bibitem{IOAKOK}
P. Ioannou and P Kokotovic, { Instability analysis and improvement of robustness of adaptive control}, {\em Automatica}, vol. 20, no. 5, pp. 583-594, 1984.

\bibitem{IOASUNbook}
P. Ioannou and J. Sun, {\em Robust Adaptive Control}, Prentice-Hall, New Jersey, 1996.

\bibitem{JIAWAN}
Z. P. Jiang and Y. Wang, Input-to-state stability for discrete-time nonlinear systems, \AUT, vol. 37, pp. 857-869, 2001.

\bibitem{KELDOW}
C. M. Kellett and P. M. Dower. Input-to-state stability, integral input-to-state stability, and $\call_2$-gain oroperties: Qualitative equivalences and interconnected systems, {\em IEEE Trans. on Aut. Control}, vol. 61, no. 1, pp. 3-17, 2016.


\bibitem{KHAbook}
H. K. Khalil, {\em Nonlinear Systems},  Third Edition, Prentice Hall, 2002.

\bibitem{KHAORT}
P. Khargonekar and R. Ortega, Comments on the robust stability analysis of adaptive controllers using normalizations, {\em IEEE Trans. on Aut. Control}, vol. 34, no. 4, pp. 478-479, 1989.


\bibitem{KORetal}
M. Korotina, J. G. Romero, S. Aranovskiy, A. Bobtsov and R. Ortega, Persistent excitation is unnecessary for on-line exponential parameter estimation: a new algorithm that overcomes this obstacle, \SCL, (submitted). {\tt (https://arxiv.org/abs/2106.08773)}.

\bibitem{KRAKHA}
J. Krause and P. Khargonekar, Parameter information content of measurable signals in direct adaptive control, {\em IEEE Trans. on Automatic Control}, vol. 32, no. 9, pp. 802-810. 1987.

\bibitem{KRE}
G. Kreisselmeier, Adaptive observers with exponential rate of convergence, \TAC, vol. 22, no. 1, pp. 2-8, 1977.

\bibitem{KRERIE}
G. Kreisselmeier and G. Rietze-Augst, Richness and excitation on an interval---with application to continuous-time adaptive control, \TAC, vol. 35, no. 2, pp. 165-171, 1990.

\bibitem{LEWetal}
F. Lewis, D. Vrabie, and K. Vamvoudakis, Reinforcement learning and feedback control: Using natural decision methods to design optimal adaptive controllers,  \CSM, vol. 32, no. 6, pp. 76-105, 2012.

\bibitem{LIO}
P.M. Lion, Rapid identification of linear and nonlinear systems, {\em AIAA Journal}, vol. 5, pp. 1835-1842, 1967.

\bibitem{LJUbook}
L. Ljung, {\em System Identification: Theory for the User}, Prentice Hall, New Jersey, 1987.

\bibitem{MORijacsp}
A. S. Morse, A comparative study of normalized and unnormalized tuning errors in parameter adaptive control, {\em Int. J.
Adaptive Control and Signal Processing}, vol. 6, pp. 309-318, 1992.

\bibitem{NARANNbook}
K. Narendra and A. Annaswamy, {\em Stable Adaptive Systems}, Prentice-Hall, New Jersey, 1989.

\bibitem{NUS}
R. Nussbaum, Some remarks on a conjecture in parameter adaptive control,  \SCL, vol. 3, pp. 243-246, 1983.

\bibitem{ORTproieee}
R. Ortega,  An on-line least-squares parameter estimator with finite convergence time, {\it Proc. IEEE}, vol. 76, no. 7, 1988.

\bibitem{ORTPRALAN}
R. Ortega, L. Praly and I. Landau, Robustness of discrete-time direct adaptive controllers, {\em IEEE Trans. on Automatic Control}, vol. 30, no. 12, pp. 1179-1187, 1985.

\bibitem{ORTLOZ}
R. Ortega and R. Lozano-Leal, A note on direct adaptive control of systems with bounded disturbances, {\it Automatica}, vol. 23, no. 2, pp.253-254, 1987.

\bibitem{ORTetalscl}
 R. Ortega, A. Bobtsov, A. Pyrkin and A. Aranovskiy, A parameter estimation approach to  state observation of nonlinear systems, {{\it Systems and Control Letters}},  vol. 85, pp 84-94, 2015.

\bibitem{ORTetal_aut19}
R. Ortega, D. Gerasimov, N. Barabanov and V. Nikiforov, Adaptive control of linear multivariable systems using dynamic regressor extension and mixing estimators: Removing the high-frequency gain assumption, \AUT, vol. 110, 108589, 2019.

\bibitem{ORTNIKGER}
R. Ortega, V. Nikiforov and D. Gerasimov, On modified parameter estimators for identification and adaptive control: a unified framework and some new schemes, \ARC, vol. 50, pp. 278-293, 2020.

\bibitem{ORTetalaut21}
R. Ortega, V. Gromov, E. Nu\~no, A. Pyrkin and J. G. Romero, Parameter estimation of nonlinearly parameterized regressions: application to system identification and adaptive control,  \AUT,  vol. 127, 109544, 2021.

\bibitem{ORTajc}
R. Ortega, Comments on recent claims about trajectories of control systems valid for particular initial conditions,  \AJC, ({\tt DOI: 10.1002/asjc.2512}), 2021.

\bibitem{ORTetaltac}
R. Ortega, S. Aranovskiy, A. Pyrkin, A Astolfi and A. Bobtsov, New results on parameter estimation via dynamic regressor extension and mixing: Continuous and discrete-time cases, \TAC, vol. 66, no. 5, pp. 2265-2272, 2021.

\bibitem{ORTetalaut}
R. Ortega, A. Bobtsov, N. Nikolayev, J. Schiffer, D. Dochain, Generalized parameter estimation-based observers: Application to power systems and chemical-biological reactors,  \AUT,  vol. 129, 109635, 2021.

\bibitem{ORTBOBNIK}
R. Ortega, A. Bobtsov and  N. Nikolayev, Parameter identification with finite-convergence time alertness preservation, \CSL, vol. 6, pp. 205-210, 2022

\bibitem{PANYU}
Y. Pan and H. Yu, Composite learning robot control with guaranteed parameter convergence, {\em Automatica}, vol. 89, pp. 398-406, 2018.

\bibitem{PANetal}
Y. Pan, S. Aranovskiy, A. Bobtsov, and H. Yu, Efficient learning from adaptive control under sufficient excitation, {\em International Journal of Robust and Nonlinear Control}, vol. 29, pp. 3111-3124, 2019.

\bibitem{PAVetal}
A. Pavlov, A. Pogromsky, N. van de Wouw and H. Nijmeijer, Convergence dynamics, a tribute to Boris Pavlovich Demidovich, \SCL, vol. 52, pp. 257-261, 2004.

\bibitem{PRAyale}
L. Praly, Robustness of model reference adaptive control, in {\em Proc. 3rd Yale Workshop on Adaptive Control}, New Haven, CT, June 15-17, 1983.

\bibitem{PRA}
L. Praly, Convergence of the gradient algorithm for linear regression models in the continuous and discrete-time cases, {\em Int. Rep. MINES ParisTech, Centre Automatique et Syst\`{e}mes}, {\tt hal.archives-ouvertes.fr/hal-01423048}, 2017.


\bibitem{ROHetal}
C. Rohrs, L. Valavani, M. Athans  and G. Stein,  Robustness of continuous-time adaptive control algorithms in the presence of unmodeled dynamics, {\em IEEE Trans. on Automatic Control}, vol. 30, no. 9, pp. 881-889, 1985.

\bibitem{RUGbook}
W.J. Rugh, {\em Linear Systems Theory}, 2nd Edition, Prentice hall, NJ, 1996.

\bibitem{SASBODbook}
S. Sastry and M. Bodson, {\em Adaptive Control: Stability, Convergence and Robustness}, Prentice-Hall, New Jersey, 1989.

\bibitem{TAObook}
G. Tao, {\em Adaptive Control Design and Analysis}, John Wiley \& Sons, New Jersey, 2003.

\bibitem{TAO}
G. Tao, Adaptive control of multivariable systems: a survey, {\em Automatica}, vol. 50, no. 11, pp. 2737-2764, 2014.

\bibitem{TYUetal}
I. Y. Tyukin, D. V. Prokhorov and C. V. Leeuwen, Adaptation and parameter estimation in systems with unstable target dynamics and nonlinear parameterization, {\it IEEE Trans. Automatic Control}, vol. 52, no. 9, pp. 1543-1559, 2007.

\bibitem{WUetal}
Z. Wu, M. Ma, X. Xu, B. Liu and Z. Yu. Predefined-time parameter estimation via modified dynamic regressor extension and mixing, {\em Journal of the Franklin Institute}, {\tt DOI: https://doi.org/10.1016/j.jfranklin.2021.06.028}, 2021.

\bibitem{YIetalcdc21}
B. Yi, C. Jin, L. Wang, G. Shi and I. R. Manchester, An almost globally convergent observer for visual SLAM without
persistent excitation, {\em 60th IEEE Conference on Decision and Control},  Austin, Texas, USA, December 13-15, 2021.

\bibitem{YIetal}
B. Yi, R. Ortega, D. Wu and W. Zhang, Orbital stabilization of nonlinear systems via Mexican sombrero energy pumping-and-damping injection. {\it Automatica}, vol. 112, 108-861, 2020.


\end{thebibliography}
\end{document}